\documentclass[fleqn,10pt]{wlscirep}
\usepackage[T1]{fontenc}
\usepackage[T1]{fontenc}
\title{Quasi-one-dimensional motion of an active MXene sheet driven by chemo-hydrodynamic waves}

\author[1,+]{Hui Wang}
\author[1,+]{Huan Liu}
\author[1]{Ling Yuan}
\author[1]{Zihao Liu}

\author[1]{Meng Zhang}

\author[2,*]{Irving R Epstein}
\author[1,*]{Qingyu Gao}
\affil[1]{China University of Mining and Technology, College of Chemical Engineering, Jiangsu, 22116 Xuzhou, P.R. China}

\affil[2]{Brandeis University,Department of Chemistry and Volen Center for Complex Systems, Waltham, 02454-9110 Massachusetts, United States}

\affil[1,*]{corresponding.author gaoqy@cumt.edu.cn}
\affil[2,*]{corresponding.author epstein@brandeis.edu}

\affil[+]{these authors contributed equally to this work}

\begin{abstract}
Signal-driven motion is widespread in natural and artificial systems, yet quantitative characterization of how transient chemo-hydrodynamic waves are converted into mechanical driving forces remains limited. Here, we investigate the self-propulsion of a MXene sheet asymmetrically coated with catalase in hydrogen peroxide solution. By combining dual-view particle image velocimetry experiments and numerical simulations reveal that active motion of the sheet is driven by chemo-hydrodynamic waves and exhibits direct-wave motion, the driving force of which is analyzed in terms of the shear stress on the sheet surface caused by chemo-hydrodynamic waves. This work suggests theoretical principles for designing and controlling hydrodynamically driven active motion.
\end{abstract}
\begin{document}

\flushbottom
\maketitle
%
%
\thispagestyle{empty}

\section*{Introduction}

The faithful transmission and processing of chemical and biochemical signals into optical, electrical, thermal and mechanical signals, and vice versa\cite{RN1}, are fundamental to life, governing essential biological processes like growth, survival, and reproduction\cite{RN2}. Inspired by nature, the field of active matter seeks to replicate this sophisticated responsiveness\cite{RN3,RN4}. The ability of active matter to perceive and transduce signals into directional motion through mechanical or adhesive force, known as signal-driven active motion, is a ubiquitous feature of artificial biomimetic systems\cite{RN5,RN6}. This motion fundamentally relies on symmetry breaking and nonreciprocal interactions engineered into the system's response. Canonical examples include propulsion by nonreciprocal push-pull force asymmetry in Belousov-Zhabotinsky waves\cite{RN5,7,8}, actuation by scanning light patterns in liquid crystals\cite{RN9}, and locomotion under programmed electric or magnetic signals\cite{RN10,RN11}.

Recent years have seen growing interest in active actuation within complex fluidic environments, owing to their adaptability and responsiveness\cite{RN12,RN13,RN14}. Previous research has demonstrated how self-generated gradients\cite{RN13,RN14} or bubbles\cite{RN15,RN16} can power active motion. For instance, studies have utilized surface tension variations induced by the dissolution\cite{RN17,RN18,RN19,RN20}, reaction\cite{RN21,RN22,RN23,RN24}, or physical interaction\cite{RN25,RN26,RN27} of surfactants to drive nonlinear velocity oscillation, translation, and rotation. However, as surfactants become depleted, such propulsion mechanisms typically exhibit poor self-sustaining ability and limited adaptability. Moreover, they often lack biocompatibility—as exemplified by systems like camphor boats\cite{RN28} or those involving organic solvents\cite{RN29}—which further limits their applicability in biologically relevant contexts. Alternatively, enzyme-catalyzed reactions have been used as a route to designing complex shape changes in flexible structures through self-induced density variations\cite{RN30,RN31}. While this actuation approach has advanced rapidly in numerical simulations, experimental methods for dynamically monitoring the coupling among flow fields, structural deformation, and directed motion still require further development. Moreover, despite the richness of observed individual motion patterns, a fundamental physical understanding of the coupling between the moving entity and the hydrodynamic waves remains elusive. In particular, the central unanswered question remains how the chemo-hydrodynamic signals determine the selection and dynamics of these motile modes. In particular, it has been shown that asymmetrically enzyme-coated sheets moving at the air–liquid interface can achieve directional propulsion driven by chemically generated counter-rotating convective vortices beneath the sheet, as demonstrated by Song and co-workers\cite{RN32}. These studies have elucidated the vortex-based propulsion mechanism and highlighted the role of chemo-hydrodynamic coupling. Despite this mechanistic understanding, quantitative characterization of the hydrodynamic driving forces remains limited. Most existing studies rely on qualitative flow visualization or steady-state analysis\cite{RN32,RN33}, and the time-resolved shear stress acting on the moving body has rarely been directly quantified. Consequently, the dynamic interaction between transient chemo-hydrodynamic waves and the mechanical response of the moving entity remains insufficiently explored.

Enzyme-coated nanomotors exhibit a wide variety of propulsion behaviors in fluid environments, offering a diverse set of models for macroscale fluid-driven actuators\cite{RN34,RN35}. The high efficiency and biocompatibility of enzymes further underpin their potential in motility-related functions and biomedical applications. A commonly employed strategy involves asymmetric structures\cite{RN36,RN37}, which generate propulsive forces through mechanisms such as self-electrophoresis and self-diffusiophoresis\cite{RN38}. In particular, enzyme-powered nanomotors—exemplified by catalase-conjugated systems—have attracted growing interest for biomedical applications\cite{RN39,RN40}. Inspired by studies on enzyme-driven nanomotors, we explore here macroscopic directional motion using a MXene\cite{RN41} sheet asymmetrically coated with catalase. We seek to uncover the origin of the driving force in a chemically active system by combining synchronous measurements of individual motion and hydrodynamic signals with numerical simulations. Through coordinated experiments and simulations, we demonstrate that the propulsion arises from self-generated chemo-hydrodynamic signals and generates directional motion. This study not only deepens the understanding of autonomous propulsion mechanisms in active systems but also establishes a foundation for the design and control of hydrodynamically driven active motion. This study advances a force-based understanding of signal-driven propulsion in active systems and provides a quantitative framework for the analysis and control of hydrodynamically driven active motion.

\section*{Results}

\textbf{Linear locomotion and dual-view particle image velocimetry (PIV) system}

In our experimental system, active asymmetric MXene sheets of composition Ti$_3$C$_2$T$_X$, with $T_X = \mathrm{-O}$, $\mathrm{-OH}$, or $\mathrm{-F}$, were fabricated via the vacuum filtration method~\cite{RN41}. Employing the conventional MXene etching route, one half of the sheet (2\,mm wide $\times$ 5\,mm long, $\sim$5\,$\mu$m thick) was modified with catalase enzyme using a mask, while leaving the other half unmodified (see Supporting Information, section A). XRD confirmed the crystal structure, and FTIR spectroscopy verified the successful covalent grafting of catalase onto MXene (Figures S2–S4), providing a reliable material basis for the subsequent active locomotion. The active sheet was placed at the air–liquid interface of a 10\,mM hydrogen peroxide solution inside a $76 \times 76 \times 5$\,mm$^3$ chamber. The large chamber dimensions minimize boundary effects from wall surface tension. Owing to its small thickness and large surface area, the sheet remains stably suspended at the interface under the balance of buoyancy and gravity.

Although linear locomotion driven by asymmetric designs has been reported in both Marangoni\cite{RN19}, bubbles\cite{RN42} and solutal buoyancy\cite{RN32} systems, here we show that radially symmetric shapes (triangles and squares) with half-side enzyme modification can also produce similar straight trajectories (Figure 1). Specifically, MXene sheets with half-side catalase modification (yellow region in Figure 1a) exhibit straight motion at the air–liquid interface of 20 mM H$_2$O$_2$, and both triangular and square designs yield the same behavior. The sheet velocity increases with substrate concentration (Figure S7) and with the liquid layer thickness, indicating solutal buoyancy rather than the Marangoni effect as the driving mechanism; in Marangoni-driven systems, the velocity is largely insensitive to the liquid thickness\cite{RN21}, and the surface tension change upon reaction is negligible (Figure S6a). We then coupled the Navier–Stokes equations with the reaction–diffusion process (detailed below and in the Supporting Information) in a finite-element model, and the computed trajectories agree well with experiments (Figure 1b). The catalytic reaction generates a solutal buoyancy–induced flow structure that drives the linear propulsion of the sheet, independent of variations in surface geometry. Experimentally, the trajectory is not perfectly straight but exhibits slight curvature. This deviation arises from nonlinear hydrodynamic effects associated with the finite width of the sheet, which modifies the surrounding flow field, as well as from unavoidable fabrication imperfections that introduce small tangential force components. Although the experiments are performed in a fully three-dimensional environment, the asymmetric structural design constrains the net motion predominantly along a single spatial direction. For clarity, we therefore describe this system as a quasi-one-dimensional (quasi-1D) system. We now turn to the critical questions: does this flow structure function analogously to a neural signal driving muscle contraction? How does fluid–structure coupling actually occur, how does the fluid signal relate to the active motion, and how can the driving force be quantified?

To fully capture this dynamic and hydrodynamic signal waves, we employed a synchronized dual-view PIV system to record the motion of the MXene sheet and the surrounding flow field from both top and side views, as illustrated in Figure 2a. The side-view camera (CCD1) performs PIV based on a standard cross-correlation algorithm: the displacement field is extracted from the statistical correlation of particle patterns in two consecutive images, and the velocity field is then computed. Simultaneously, the top-view camera (CCD2) tracks the motion of the active sheet, enabling precise correlation between the flow-field signals and the active motion (detailed setup in the Experiment and Simulation section; see Figure S5 for the apparatus). With this synchronized acquisition, we obtain concurrent information on the active sheet and the flow field, as shown in Figure 1b. The side-view (CCD1) results reveal two convective vortices generated by solutal buoyancy from the decomposition of hydrogen peroxide by the active sheet, synchronized with its motion (see Video S1). The top-view (CCD2) results show the sheet translating linearly to the left; white dots on its left and right sides mark the boundaries of the side-view field of view, facilitating subsequent spatiotemporal alignment. Because the microscopic flow field must be resolved, the side-view field of view is much smaller than the top-view one—a design that overcomes the challenge of simultaneously capturing macroscopic motion and microscopic flow information.

\textbf{Chemo-hydrodynamic Signals}

In complex fluidic environments, diverse signals—such as pressure, temperature, and flow velocity—are ubiquitous. Given the closed configuration of our experimental system and the negligible heat generation\cite{RN43}, we focus on the real-time flow velocity signal generated by fluid motion driven by the reaction-induced density variations. Specifically, by examining the flow speed distribution along the horizontal plane adjacent to the sheet (as depicted in Figure 3a,3b), we observed that the speed signal propagates in the form of a wave synchronized with the motion of the active sheet along the x-direction. Figure 3c presents the time-resolved flow profiles in the horizontal plane at the level of the lower boundary of the active sheet. To quantify the signal propagation, we analyze the absolute value of the $x$-component of the velocity field ($|u_x|$), which highlights the magnitude of the propagating velocity maxima independent of flow direction and facilitates tracking their spatial evolution. This representation emphasizes the propagating intensity envelope, while preserving directional information in the supplementary signed velocity maps. The signed x-component and the full vector field are provided in Figure S9, where two extrema of opposite sign are clearly observed, corresponding to a pair of counter-rotating vortices. The chosen horizontal plane is representative, as velocity profiles extracted at other vertical positions (Figures S9a and S9b) exhibit similar spatial distributions and peak locations, indicating that the propagating structure is robust across the vertical section. The active sheet moves from the enzyme-coated side toward the bare side, defined here as the negative x-direction (i.e., leftward). Here, the term "wave" does not refer to random fluctuations in the global centroid velocity. Instead, it describes structured, propagating features in the spatial velocity field that travel along the sheet.

To visualize and gain deeper insight into the driving mechanism, we performed finite element-based numerical simulations of this fluid–structure interaction scenario with Arbitrary Lagrangian Eulerian meshing (see Experiment and Simulation section and Supporting Information Section B for details). Using the characteristic velocity $u = 30\ \mu\mathrm{m\,s^{-1}}$, characteristic length $L = 5\ \mathrm{mm}$, density $\rho \approx 1\ \mathrm{g\,cm^{-3}}$, diffusion coefficient $D \approx 1 \times 10^{-9}\ \mathrm{m^2\,s^{-1}}$, and dynamic viscosity $\mu \approx 1 \times 10^{-3}\ \mathrm{Pa\cdot s}$, the relevant dimensionless numbers were estimated. The Reynolds number is estimated as: $Re = \rho u L / \mu \approx 0.15$, indicating a laminar, viscous-dominated flow regime. The P\'{e}clet number is estimated as: $Pe = u L / D \approx 150$, suggesting that advective transport is significant compared to diffusion, while molecular diffusion remains non-negligible. The solutal Rayleigh number is estimated as: $Ra = g \beta \Delta c L^3 / (v D) \approx 9 \times 10^4$, where concentration $\Delta c = 0.01\ \mathrm{M}$, kinematic viscosity $v = \mu / \rho \approx 10^{-6}\ \mathrm{m^2\,s^{-1}}$, solvent expansion coefficient $\beta = 0.01056\ \mathrm{M^{-1}}$~\cite{RN30}, which is larger than the critical value ($\sim 1700$), indicating a strongly buoyancy-driven unstable regime. These results confirm that the system operates in a low-Reynolds-number chemo-hydrodynamic regime, where viscous forces dominate inertia, and reaction-driven density gradients induce convection coupled with diffusion.

The simulations did not account for reaction-induced surface tension changes, given their negligible magnitude (Figure S6a), and excluded bubble effects, as no significant bubble detachment or rupture was observed via CCD monitoring in the side views (Video S1). As shown in Figure 3d, the simulated chemo-hydrodynamic signal closely resembles the experimentally measured wave signal in morphology. Both experiment and simulation show that the chemo-hydrodynamic signal around the sheet exhibits a two-peak structure—a larger peak followed by a smaller one—which corresponds to two separate convection rolls located on opposite sides of the catalytic region. Although the relative height of the two peaks varies at different vertical positions, the absolute velocity of the larger peak consistently dominates in the direction of motion. This indicates that the larger peak dictates the propulsion direction, a result of fluid-solid interaction, the asymmetric catalytic coating, and spatial flow.

\textbf{Direct wave locomotion}

The driving of the active sheet by fluidic wave signals, resulting in unidirectional motion characterized by direct wave propagation, exhibits similarities to the direct wave locomotion observed in snails\cite{RN44}. As illustrated in Figure 4, we characterize this chemo-hydrodynamic direct wave behavior of the active sheet through combined experimental and numerical approaches (see Videos S2 and S3). Experimentally, both the centroid of the sheet (black curve) and the peak position of the adjacent fluidic signal (red curve) propagate in the negative direction (Figure 4a). This behavior is reproduced in the numerical simulations (Figure 4c). Notably, the temporal evolution of this process is correlated with the variation in the sheet velocity (Figures 4b and 4d). Due to experimental noise and instabilities at the air–liquid interface, the velocity fluctuations measured in experiments are less regular than those obtained under idealized simulation conditions. Nevertheless, Fourier analysis (Figure S11c) of the experimental velocity signal reveals a dominant low-frequency component, indicating that the motion is non-uniform in time and exhibits periodic modulation rather than constant-speed translation. Together, these results support the conclusion that the chemo-hydrodynamic wave sustains the directional motion of the active sheet, consistent with a direct wave-driven propulsion mechanism characterized by cyclic modulation and resetting.

This resetting originates from the coupled dynamics between the sheet motion and the replenishment of reactant concentration in newly accessed regions through motion-induced feedback. Catalytic reactions occurring on the sheet surface generate transient fluid waves that drive directional propulsion. As the sheet advances into fresh solution, renewed reactant supply initiates subsequent waves, establishing a self-sustained feedback loop between the chemo-hydrodynamic signaling and locomotion. The periodic regeneration of these waves results in oscillations in the propulsion speed. Velocity oscillations in fluidic active systems have been widely reported\cite{RN12,RN13}. For example, oscillatory dynamics may arise from the competition between Marangoni stresses and solutal buoyancy, which involves spatiotemporal coupling\cite{RN45}. The speed variation observed in the present system is analogous to other active systems, such as Marangoni swimmers\cite{RN21} and light-powered rotating rings\cite{RN46,RN47}, where periodic regeneration of asymmetric driving forces sustains directional motion. In our case, the coupling between reaction-induced density gradients and nonlinear convective flow constitutes the intrinsic driving mechanism of this self-propelled active system.

\textbf{Driving force analysis}

To better visualize how the chemical fluid signal drives the active carrier, we directly computed the net driving force from both experiments and simulations. The active MXene sheet operates at a horizontal air–liquid interface, and its thickness is much smaller than its lateral dimensions. Because the sheet is geometrically symmetric about the axis of motion and floats on the water surface, the horizontal components of the pressure forces integrated over the sheet cancel out. Consequently, the net fluidic driving force along the direction of motion is provided exclusively by the viscous shear stress from the water side. The hydrodynamic force acting on the active sheet is obtained by integrating the viscous shear stress over the fluid–solid contact area:
\begin{equation}
\tau_{zx} = \mu \left( \frac{\partial u_x}{\partial z} + \frac{\partial u_z}{\partial x} \right)
\label{eq:shear}
\end{equation}
where $u_x$ and $u_z$ are the velocity components in the $x$- and $z$-directions, and $\mu$ is the dynamic viscosity. Based on boundary-layer theory, $\partial u_z / \partial x$ is neglected, and $\partial u_x / \partial z$ is approximated from the $x$-velocity at different $z$ heights using a forward-difference scheme. For equally spaced points ($z_1 = 4.8\ \mathrm{mm}$, $z_2 = 4.6\ \mathrm{mm}$), the second-order forward-difference formula for the velocity gradient at $z = 5\ \mathrm{mm}$ is:
\begin{equation}
\left. \frac{\partial u}{\partial z} \right|_{z = 5\ \mathrm{mm}} \approx \frac{-3u_0 + 4u_1 - u_2}{2h}
\label{eq:gradient}
\end{equation}
where $u_0$ is the sheet center-of-mass velocity (the $x$-velocity at $z = 5\ \mathrm{mm}$), and $u_1, u_2$ are the $x$-velocity components at $z_1$ and $z_2$, with $h = 0.2\ \mathrm{mm}$. The total force in the $x$-direction is then computed via the trapezoidal rule:
\begin{equation}
F_x = \iint_A \tau_{zx} \, \mathrm{d}A = W \int_{x_\mathrm{l}}^{x_\mathrm{r}} \tau_{zx}(x) \, \mathrm{d}x \approx W \cdot \frac{\Delta x}{2} \left[ \tau_{zx}(x_1) + 2\sum_{i=2}^{n-1} \tau_{zx}(x_i) + \tau_{zx}(x_n) \right]
\label{eq:force}
\end{equation}
Here, $W$ is the sheet width, $L = x_\mathrm{r} - x_\mathrm{l}$ is the sheet length, and $A$ is the wetted fluid–solid interface. The instantaneous net force evaluated from the experiments is plotted together with the signal in Figure~5a. The corresponding force obtained from the numerical simulation is shown as the red curve in Figure~5b; because the simulation employs a simplified two-dimensional model that omits the $y$-axis dimension, the simulated force represents the $x$-direction force per unit width of the sheet. Comparing the two, the experimental force exhibits a weaker local correlation with the signal variation than the simulation, owing to environmental fluctuations and limited measurement precision, yet the magnitudes are of the similar order.

From the numerical simulation, this temporal variation in force correlates with changes in the signal velocity (Figure~5a, black curve; calculated from the displacement in Figure~4c), where the decrease in net propulsion induces a drop in the speeds of the fluid signal and sheet displacement at periodic signal resetting. Due to the asymmetric catalytic patterning of the sheet, the peak of the fluid velocity signal is consistently localized to the left of the sheet's centroid (Figure~S13c). This asymmetric signal distribution and its temporal evolution induce corresponding differential changes in surface traction, which originates from the viscous shear stress at the fluid–solid interface (modeled as a boundary condition in the 3D simulation; Figure~S12). In contrast, when the signal distribution is symmetric—i.e., when the active MXene sheet is fully coated, the sheet exhibits almost no motion, as shown in Figure~S9 of the Supporting Information. The modulation of the signal velocity, in turn, stems from the dynamic concentration field, where local H$_2$O$_2$ is depleted by catalysis and then replenished convectively as the sheet moves forward into fresh solution regions (detailed in Figure~S13a,b). Thus, the reaction sustains a propagating fluid signal that acts on the sheet to produce propulsion. This motion, in turn, brings fresh catalyst regions into contact with new substrate, thereby establishing a continuous cycle of signal generation and directional locomotion.

\section*{Discussion}

To conclude, we have experimentally and numerically verified that the driving mode of the active sheet is governed by chemo-hydrodynamic signals. The fluid signal propagates in the same direction as the motion of the active sheet, exhibiting characteristics of direct wave locomotion. The movement of the catalytic zone induces variations in the flow velocity signal, and this signal, through interaction with the active sheet, generates a viscous shear force that drives the sheet’s motion. The asymmetric wave signal is identified as the key factor enabling directional motion. The active motion is rectilinear, with the feedback from movement into fresh regions and fluid nonlinearity resulting in oscillatory speed dynamics, manifesting spatially as a wave signal. Further mechanical analysis confirms the interdependent relationship between the chemo-hydrodynamic signal and the driven motion, which is essential for sustaining the continuous propulsion cycle.

While the current system requires further improvement in fluid monitoring accuracy, the active sheet—composed of biocompatible materials—offers considerable programmability in motion control. This includes the flexibility of incorporating diverse catalytic areas, types of enzymes that result in different density changes, and substrate concentration. Integrating multiple such actuators may lead to emergent collective motion driven by chemo-hydrodynamic signaling. Moreover, it holds promising potential as a drug delivery platform, as well as for multifunctional applications such as photothermal\cite{RN48} and pharmacological therapies\cite{RN49}. In a broader context, given recent advances in mechanical signaling during embryonic development\cite{RN50,RN51}, we anticipate that chemo-hydrodynamic signals will attract growing attention in future research on cell fate specification.

\section*{Methods}

\section{Materials}
Ti$_3$AlC$_2$ powder (400 mesh) was purchased from Jilin 11 Technology Co., Ltd. Lithium fluoride (LiF, 99\%), (3-aminopropyl)triethoxysilane (APTES, 99\%), polyamide resin particle (10\,$\mu$m), glutaraldehyde (25\% in water), poly(vinyl alcohol) (PVA, $M_w = 89000$), and catalase (EC1.11.1.6, from bovine liver) were supplied by Shanghai Aladdin Biochemical Technology Co. Ethanol (EtOH, AR), hydrochloric acid (HCl, 36\%), hydrogen peroxide (H$_2$O$_2$, 30\%), acetic acid (30\%), and ammonium hydroxide (NH$_3\cdot$H$_2$O, AR) were obtained from Sinopharm Chemical Reagent Co. (China). Hydrogen peroxide was used without purification and titrated with standard potassium permanganate. All reagents were used as received.

\section{Synthesis of MXene nanosheets}
The MXene nanosheets were synthesized according to a previously reported method with modifications~\cite{RN52}. Briefly, 2\,g of lithium fluoride was thoroughly mixed with 40\,mL of 9\,M hydrochloric acid in a 100\,mL polytetrafluoroethylene (PTFE) beaker under continuous stirring. Subsequently, 2\,g of Ti$_3$AlC$_2$ powder was gradually introduced into the mixture, which was then maintained at 35\,$^\circ$C with constant agitation for 24\,h. The resultant suspension underwent centrifugation at 3500\,rpm, followed by repeated washing with deionized water until the supernatant reached a pH value above 6. The obtained black sediment was redispersed in deionized water and subjected to ultrasonication for 1\,h to achieve exfoliation. After subsequent centrifugation at 3500\,rpm for 1\,h, the dark colloidal supernatant containing few-layer Ti$_3$C$_2$T$_x$ nanosheets was carefully collected. The dispersion was subjected to further centrifugation at 12000\,rpm to collect the nanosheets.

\section{Synthesis of amino-functionalized \texorpdfstring{Ti$_3$C$_2$T$_x$}{Ti3C2Tx} (\texorpdfstring{Ti$_3$C$_2$T$_x$-NH$_2$}{Ti3C2Tx-NH2})}
Ti$_3$C$_2$T$_x$-NH$_2$ nanosheets were synthesized via sol-gel condensation~\cite{RN53}. Briefly, APTES (1.0\,mL) was added to a water/ethanol solution (19:1 V/V, 20\,mL), and the pH was adjusted to 4.5 using acetic acid (1\,M). The solution was stirred at 25\,$^\circ$C for 10\,h. Subsequently, Ti$_3$C$_2$T$_x$ (100\,mg) was added, and the pH was adjusted to 8.5 with NH$_3\cdot$H$_2$O. The dispersion was refluxed at 70\,$^\circ$C for 6\,h. The product was collected by centrifugation (11,000\,rpm, 10\,min), washed with ethanol/deionized water (4:1 V/V), redispersed in deionized water, and sonicated in an ice bath for 1\,h.

\section{Fabrication of composite MXene/Enzyme sheet}
The asymmetric catalytic MXene sheet was fabricated using a vacuum-assisted filtration strategy combined with a mask-guided secondary deposition approach. Specifically, 0.1\,wt\% polyvinyl alcohol (PVA) solution (PVA was incorporated to enhance mechanical strength, aqueous stability, and enzyme compatibility~\cite{RN54}), 1\,mg\,mL$^{-1}$ catalase, and 100\,$\mu$L glutaraldehyde (serving as crosslinker) were sequentially introduced into 5\,mg\,mL$^{-1}$ Ti$_3$C$_2$T$_x$--NH$_2$ colloidal suspension under continuous magnetic stirring for 1\,h. The resulting mixture was then subjected to a two-step filtration process through a 22\,$\mu$m cellulose membrane acting as a patterned mask. The primary filtration deposited the functional composite layer, followed by secondary deposition with pure 5\,mg\,mL$^{-1}$ MXene solution to establish an asymmetric distribution. The obtained sheet was subsequently dried at 4\,$^\circ$C for 12\,h before being carefully peeled off from the membrane substrate (Figure~S1).

\section{Actuation device and characterization}
The experiments were performed in a square container (76\,mm $\times$ 76\,mm $\times$ 5\,mm) filled with an aqueous H$_2$O$_2$ solution. Unless otherwise specified, all solute concentrations were 10\,mM, and the solution thickness was 5\,mm. The sheet motion was recorded by two synchronized CCD cameras at 5\,fps: one acquired top-view (XY) trajectories, and the other—equipped with a zoom lens (0.1X--1.8X)—recorded side-view (XZ) profiles for flow field analysis, with its focus set on the rectilinear motion in the XZ plane while simultaneously monitoring XY displacements. An LED area light (0.5\,mW\,cm$^{-2}$) was used for side illumination to enhance particle contrast. This configuration also provided background illumination for the top-view camera, thus avoiding the use of lasers and preventing photothermal effects. Trajectory data and sheet speed were processed with Fiji (v2.14.0), and velocity fields were derived using PIVlab software (v3.08). A standard H$_2$O$_2$ concentration of 10\,mM was used, and the sheet samples were characterized by SEM, XRD, and AFM (detailed characterization information can be found in Section A of the supporting information).

\section{Simulation}
To simulate the motion driven by the interaction between catalytic sheets and hydrodynamic effects, numerical simulations were performed using the diffusion and fluid-structure interaction modules in COMSOL Multiphysics. The computational domain is a 76\,mm $\times$ 76\,mm $\times$ 5\,mm rectangular prism chamber. A sheet measuring 2\,mm $\times$ 5\,mm and 0.1\,mm in thickness is placed at the top boundary, representing a sheet floating at the liquid-air interface. The catalytic region is located in the right half of the sheet. The model neglects fluid flow induced by reaction exothermicity~\cite{RN43}. Meanwhile, the Young's modulus of the MXene sheet is set to be relatively high (1\,GPa), and its elastic deformation is neglected. The reaction-diffusion system is governed by the following flux equations\cite{RN32}:
\begin{equation}
\frac{\partial c_i}{\partial t} + \nabla \cdot \mathbf{J}_i + \mathbf{u} \cdot \nabla c_i = R_i
\label{eq:reaction_diffusion}
\end{equation}
\begin{equation}
\mathbf{J}_i = -D_i \nabla c_i
\label{eq:flux}
\end{equation}
where $\mathbf{J}_i$ represents the flux of species $i$, whose concentration is $c_i$, $D_i$ denotes the corresponding diffusion coefficient, and $R_i$ accounts for chemical production/consumption rates during reactions.

The enzymatic reaction kinetics follow the Michaelis-Menten relationship:
\begin{equation}
R = \frac{R_{\max} [C]}{K_m + [C]}
\label{eq:michaelis_menten}
\end{equation}
where $R_{\max}$ represents the maximum reaction rate defined as:
\begin{equation}
R_{\max} = w k_{\mathrm{cat}} [E]
\label{eq:Rmax}
\end{equation}
Here, $K_m$ corresponds to the Michaelis constant (substrate concentration at half-maximal reaction rate), $C$ is the substrate concentration, $w$ denotes the number of active enzyme sites ($w = 4$), $k_{\mathrm{cat}}$ signifies the catalytic turnover number, and $[E]$ represents the surface enzyme concentration.

Fluid dynamics were solved using the Navier-Stokes equation:
\begin{equation}
\rho \left( \frac{\partial \mathbf{u}}{\partial t} + \mathbf{u} \cdot \nabla \mathbf{u} \right) = \nabla \cdot \left[ -p \mathbf{I} + \mu \left( \nabla \mathbf{u} + (\nabla \mathbf{u})^T \right) \right] + \mathbf{F}_b
\label{eq:navier_stokes}
\end{equation}
with the incompressibility constraint:
\begin{equation}
\nabla \cdot \mathbf{u} = 0
\label{eq:incompressibility}
\end{equation}
and
\begin{equation}
\mathbf{F}_b = -g \Delta \rho = \rho_0 \sum_{i}^{j} \beta_i C_i
\label{eq:buoyancy}
\end{equation}
where $p$ denotes pressure, $\rho_0$ is the solvent density, $\rho$ is the real-time fluid density, $\mathbf{I}$ is the identity tensor, $\mu$ is the dynamic viscosity, $\mathbf{u}$ is the fluid velocity, $\mathbf{F}_b$ is the buoyancy force, and $\beta$ is the expansion coefficient~\cite{RN55}. The term $(\nabla \mathbf{u})^T$ is the transpose of the fluid velocity gradient. Other parameter settings can be found in Section B of the supplementary material.

\bibliography{sample}

@article{RN1,
   author = {Stuart, Martien A. Cohen and Huck, Wilhelm T. S. and Genzer, Jan and Müller, Marcus and Ober, Christopher and Stamm, Manfred and Sukhorukov, Gleb B. and Szleifer, Igal and Tsukruk, Vladimir V. and Urban, Marek and Winnik, Françoise and Zauscher, Stefan and Luzinov, Igor and Minko, Sergiy},
   title = {Emerging applications of stimuli-responsive polymer materials},
   journal = {Nature Materials},
   volume = {9},
   number = {2},
   pages = {101-113},
   ISSN = {1476-4660},
   DOI = {https://doi.org/10.1038/nmat2614},
   url = {https://doi.org/10.1038/nmat2614},
   year = {2010},
   type = {Journal Article}
}

@Article{RN2,
author ="Brunetti, A. E. and Carnevale Neto, F. and Vera, M. C. and Taboada, C. and Pavarini, D. P. and Bauermeister, A. and Lopes, N. P.",
title  ="An integrative omics perspective for the analysis of chemical signals in ecological interactions",
journal  ="Chem. Soc. Rev.",
year  ="2018",
volume  ="47",
issue  ="5",
pages  ="1574-1591",
publisher  ="The Royal Society of Chemistry",
doi  ="https://doi.org/10.1039/C7CS00368D",
url  ="http://dx.doi.org/10.1039/C7CS00368D",
}

@article{RN3,
   author = {Kaspar, C. and Ravoo, B. J. and van der Wiel, W. G. and Wegner, S. V. and Pernice, W. H. P.},
   title = {The rise of intelligent matter},
   journal = {Nature},
   volume = {594},
   number = {7863},
   pages = {345-355},
   ISSN = {1476-4687 (Electronic)
0028-0836 (Linking)},
   DOI = {10.1038/s41586-021-03453-y},
   url = {https://www.ncbi.nlm.nih.gov/pubmed/34135518},
   year = {2021},
   type = {Journal Article}
}

@article{RN4,
   author = {Walther, Andreas},
   title = {Viewpoint: From Responsive to Adaptive and Interactive Materials and Materials Systems: A Roadmap},
   journal = {Advanced Materials},
   volume = {32},
   number = {20},
   pages = {1905111},
   ISSN = {0935-9648},
   DOI = {https://doi.org/10.1002/adma.201905111},
   url = {https://doi.org/10.1002/adma.201905111},
   year = {2020},
   type = {Journal Article}
}

@article{RN5,
   author = {Ren, Lin and Yuan, Ling and Gao, Qingyu and Teng, Rui and Wang, Jing and Epstein, Irving R.},
   title = {Chemomechanical origin of directed locomotion driven by internal chemical signals},
   journal = {Science Advances},
   volume = {6},
   number = {18},
   
   ISSN = {2375-2548},
   DOI = {https://doi.org/10.1126/sciadv.aaz9125},
   url = {<Go to ISI>://WOS:000531089700038},
   year = {2020},
   type = {Journal Article}
}

@article{RN6,
   author = {Ghosh, Souvik and Baltussen, Mathieu G. and Ivanov, Nikita M. and Haije, Rianne and Jakštaitė, Miglė and Zhou, Tao and Huck, Wilhelm T. S.},
   title = {Exploring Emergent Properties in Enzymatic Reaction Networks: Design and Control of Dynamic Functional Systems},
   journal = {Chemical Reviews},
   volume = {124},
   number = {5},
   pages = {2553-2582},
   
   ISSN = {0009-2665},
   DOI = {10.1021/acs.chemrev.3c00681},
   url = {https://doi.org/10.1021/acs.chemrev.3c00681},
   year = {2024},
   type = {Journal Article}
}

@article{7,
author = {Lin Ren  and Meng Wang  and Changwei Pan  and Qingyu Gao  and Yang Liu  and Irving R. Epstein },
title = {Autonomous reciprocating migration of an active material},
journal = {Proceedings of the National Academy of Sciences},
volume = {114},
number = {33},
pages = {8704-8709},
year = {2017},
doi = {https://doi.org/10.1073/pnas.1704094114},
URL = {https://www.pnas.org/doi/abs/10.1073/pnas.1704094114},
}

@article{8,
author = {Yu, Haodi and Ren, Lin and Wang, Yunjie and Wang, Hui and Zhang, Meng and Pan, Changwei and Yuan, Ling and Zhang, Jiujun and Epstein, Irving R. and Gao, Qingyu},
title = {Chiral Locomotion Transitions of an Active Gel and Their Chemomechanical Origin},
journal = {Journal of the American Chemical Society},
volume = {147},
number = {6},
pages = {5182-5188},
year = {2025},
doi = {https://doi.org/10.1021/jacs.4c15550},
}

@article{RN9,
   author = {Palagi, Stefano and Mark, Andrew G and Reigh, Shang Yik and Melde, Kai and Qiu, Tian and Zeng, Hao and Parmeggiani, Camilla and Martella, Daniele and Sanchez-Castillo, Alberto and Kapernaum, Nadia},
   title = {Structured light enables biomimetic swimming and versatile locomotion of photoresponsive soft microrobots},
   journal = {Nature materials},
   volume = {15},
   number = {6},
   pages = {647-653},
   ISSN = {1476-1122},
   DOI = {https://doi.org/10.1038/nmat4569},
   year = {2016},
   type = {Journal Article}
}

@article{RN10,
   author = {Hu, Wenqi and Lum, Guo Zhan and Mastrangeli, Massimo and Sitti, Metin},
   title = {Small-scale soft-bodied robot with multimodal locomotion},
   journal = {Nature},
   volume = {554},
   number = {7690},
   pages = {81-85},
   ISSN = {0028-0836
1476-4687},
   DOI = {https://doi.org/10.1038/nature25443},
   year = {2018},
   type = {Journal Article}
}

@article{RN11,
   author = {Lee, Jin Gyun and Brooks, Ada M. and Shelton, William A. and Bishop, Kyle J. M. and Bharti, Bhuvnesh},
   title = {Directed propulsion of spherical particles along three dimensional helical trajectories},
   journal = {Nature Communications},
   volume = {10},
   number = {1},
   pages = {2575},
   ISSN = {2041-1723},
   DOI = {https://doi.org/10.1038/s41467-019-10579-1},
   url = {https://doi.org/10.1038/s41467-019-10579-1},
   year = {2019},
   type = {Journal Article}
}

@article{RN12,
   author = {Budroni, M. A. and Rossi, F.},
   title = {Transport-driven chemical oscillations: a review},
   journal = {Physical Chemistry Chemical Physics},
   volume = {26},
   number = {47},
   pages = {29185-29226},
   ISSN = {1463-9076},
   DOI = {10.1039/D4CP03466J},
   url = {http://dx.doi.org/10.1039/D4CP03466J},
   year = {2024},
   type = {Journal Article}
}

@article{RN13,
   author = {Manna, Raj Kumar and Laskar, Abhrajit and Shklyaev, Oleg E and Balazs, Anna C},
   title = {Harnessing the power of chemically active sheets in solution},
   journal = {Nature Reviews Physics},
   volume = {4},
   number = {2},
   pages = {125-137},
   ISSN = {2522-5820},
   year = {2022},
   DOI = {https://doi.org/10.1038/s42254-021-00395-2},
   type = {Journal Article}
}

@article{RN14,
   author = {Suematsu, Nobuhiko J. and Nakata, Satoshi},
   title = {Evolution of Self-Propelled Objects: From the Viewpoint of Nonlinear Science},
   journal = {Chemistry--A European Journal},
   volume = {24},
   number = {24},
   pages = {6308-6324},
  
   ISSN = {0947-6539},
   DOI = {https://doi.org/10.1002/chem.201705171},
   url = {https://chemistry-europe.onlinelibrary.wiley.com/doi/abs/10.1002/chem.201705171},
   year = {2018},
   type = {Journal Article}
}

@article{RN15,
author = {Nakata, Satoshi and Nomura, Mio and Yamamoto, Hiroya and Izumi, Shunsuke and Suematsu, Nobuhiko J. and Ikura, Yumihiko and Amemiya, Takashi},
title = {Periodic Oscillatory Motion of a Self-Propelled Motor Driven by Decomposition of H2O2 by Catalase},
journal = {Angewandte Chemie International Edition},
volume = {56},
number = {3},
pages = {861-864},
doi = {https://doi.org/10.1002/anie.201609971},
url = {https://onlinelibrary.wiley.com/doi/abs/10.1002/anie.201609971},
year = {2017}
}

@article{RN16,
   author = {Kumar, BVVS Pavan and Patil, Avinash J and Mann, Stephen},
   title = {Enzyme-powered motility in buoyant organoclay/DNA protocells},
   journal = {Nature chemistry},
   volume = {10},
   number = {11},
   pages = {1154-1163},
   ISSN = {1755-4330},
   year = {2018},
    DOI = {https://doi.org/10.1038/s42254-021-00395-2},
   type = {Journal Article}
}

@article{RN17,
   author = {Pena-Francesch, Abdon and Giltinan, Joshua and Sitti, Metin},
   title = {Multifunctional and biodegradable self-propelled protein motors},
   journal = {Nature communications},
   volume = {10},
   number = {1},
   pages = {3188},
   ISSN = {2041-1723},
   year = {2019},
  DOI = {https://doi.org/10.1038/s41467-019-11141-9}, 
   type = {Journal Article}
}

@article{RN18,
   author = {Hokmabad, Babak Vajdi and Agudo-Canalejo, Jaime and Saha, Suropriya and Golestanian, Ramin and Maass, Corinna C},
   title = {Chemotactic self-caging in active emulsions},
   journal = {Proceedings of the National Academy of Sciences},
   volume = {119},
   number = {24},
   pages = {e2122269119},
   ISSN = {0027-8424},
   year = {2022},
    DOI = {https://doi.org/10.1073/pnas.2122269119}, 
   type = {Journal Article}
}

@article{RN19,
   author = {Song, Seo Woo and Lee, Sumin and Choe, Jun Kyu and Lee, Amos Chungwon and Shin, Kyoungseob and Kang, Junwon and Kim, Gyeongjun and Yeom, Huiran and Choi, Yeongjae and Kwon, Sunghoon},
   title = {Pen-drawn Marangoni swimmer},
   journal = {Nature Communications},
   volume = {14},
   number = {1},
   pages = {3597},
   ISSN = {2041-1723},
   year = {2023},
    DOI = {https://doi.org/10.1038/s41467-023-39186-x},
   type = {Journal Article}
}

@article{RN20,
   author = {Kumar, Manoj and Murali, Aniruddh and Subramaniam, Arvin Gopal and Singh, Rajesh and Thutupalli, Shashi},
   title = {Emergent dynamics due to chemo-hydrodynamic self-interactions in active polymers},
   journal = {Nature Communications},
   volume = {15},
   number = {1},
   pages = {4903},
   ISSN = {2041-1723},
   year = {2024},
    DOI = {https://doi.org/10.1038/s41467-024-49155-7},
   type = {Journal Article}
}

@article{RN21,
   author = {Holstein, Lara Rae and Suematsu, Nobuhiko J. and Takeuchi, Masayuki and Harano, Koji and Banno, Taisuke and Takai, Atsuro},
   title = {Reduction-Induced Self-Propelled Oscillatory Motion of Perylenediimides on Water},
   journal = {Angewandte Chemie International Edition},
   volume = {63},
   number = {46},
   pages = {e202410671},
   ISSN = {1433-7851},
   DOI = {https://doi.org/10.1002/anie.202410671},
   url = {https://onlinelibrary.wiley.com/doi/abs/10.1002/anie.202410671},
   year = {2024},
   type = {Journal Article}
}

@article{RN22,
author = {Nguindjel, Anne-D{\'e}borah C. and Franssen, Stan C. M. and Korevaar, Peter A.},
title = {Reconfigurable Droplet–Droplet Communication Mediated by Photochemical Marangoni Flows},
journal = {Journal of the American Chemical Society},
volume = {146},
number = {9},
pages = {6006-6015},
year = {2024},
doi = {https://doi.org/10.1021/jacs.3c12882},

}

@article{RN23,
   author = {Jamaluddin, Syed Jazli Syed and Khaothong, Kritsana and Tinsley, Mark R and Showalter, Kenneth},
   title = {Photochemical motion control of surface active Belousov–Zhabotinsky droplets},
   journal = {Chaos: An Interdisciplinary Journal of Nonlinear Science},
   volume = {30},
   number = {8},
   pages = {083143},
   ISSN = {1054-1500},
   year = {2020},
   doi = {https://doi.org/10.1063/5.0016252},
   type = {Journal Article}
}

@article{RN24,
   author = {Zhu, Guiqiang and Zhang, Shu and Lu, Guoxin and Peng, Benwei and Lin, Cuiling and Zhang, Liqun and Shi, Feng and Zhang, Qian and Cheng, Mengjiao},
   title = {ON–OFF Control of Marangoni Self‐propulsion via A Supra‐amphiphile Fuel and Switch},
   journal = {Angewandte Chemie International Edition},
   volume = {63},
   number = {30},
   pages = {e202405287},
   ISSN = {1433-7851},
   year = {2024},
   doi = {https://doi.org/10.1002/anie.202405287},
   type = {Journal Article}
}

@article{RN25,
   author = {Hou, Kai and Guan, Dongshi and Li, Hangyu and Sun, Yongqi and Long, Yue and Song, Kai},
   title = {Programmable light-driven swimming actuators via wavelength signal switching},
   journal = {Science Advances},
   volume = {7},
   number = {37},
   pages = {eabh3051},
   ISSN = {2375-2548},
   year = {2021},
   doi = {https://doi.org/10.1126/sciadv.abh3051},
   type = {Journal Article}
}

@article{RN26,
   author = {Lin, Hai and Qian, Yongqiang and Zhou, Peidi and Lin, Jian and Luo, Zhiling and Zhang, Wei and Chen, Luzhuo},
   title = {Electricity-Driven Strategies for Bioinspired Multifunctional Swimming Marangoni Robots Based on Super-Aligned Carbon Nanotube Composites},
   journal = {Small},
   volume = {20},
   number = {33},
   pages = {2400906},
   ISSN = {1613-6810},
   DOI = {https://doi.org/10.1002/smll.202400906},
   url = {https://onlinelibrary.wiley.com/doi/abs/10.1002/smll.202400906},
   year = {2024},
   type = {Journal Article}
}

@article{RN27,
   author = {Tang, Shi-Yang and Sivan, Vijay and Petersen, Phred and Zhang, Wei and Morrison, Paul D. and Kalantar-zadeh, Kourosh and Mitchell, Arnan and Khoshmanesh, Khashayar},
   title = {Liquid Metal Actuator for Inducing Chaotic Advection},
   journal = {Advanced Functional Materials},
   volume = {24},
   number = {37},
   pages = {5851-5858},
   ISSN = {1616-301X},
   DOI = {https://doi.org/10.1002/adfm.201400689},
   url = {https://advanced.onlinelibrary.wiley.com/doi/abs/10.1002/adfm.201400689},
   year = {2014},
   type = {Journal Article}
}

@article{RN28,
   author = {Nakata, Satoshi and Iguchi, Yasutaka and Ose, Sachie and Kuboyama, Makiko and Ishii, Toshio and Yoshikawa, Kenichi},
   title = {Self-rotation of a camphor scraping on water: new insight into the old problem},
   journal = {Langmuir},
   volume = {13},
   number = {16},
   pages = {4454-4458},
   ISSN = {0743-7463},
   year = {1997},
   DOI = {https://doi.org/10.1021/la970196p},
   type = {Journal Article}
}

@article{RN29,
   author = {Dwivedi, Prateek and Pillai, Dipin and Mangal, Rahul},
   title = {Self-propelled swimming droplets},
   journal = {Current Opinion in Colloid and Interface Science},
   volume = {61},
   pages = {101614},
   ISSN = {1359-0294},
   year = {2022},
   DOI = {https://doi.org/10.1016/j.cocis.2022.101614},
   type = {Journal Article}
}

@article{RN30,
   author = {Manna, Raj Kumar and Shklyaev, Oleg E and Balazs, Anna C},
   title = {Chemical pumps and flexible sheets spontaneously form self-regulating oscillators in solution},
   journal = {Proceedings of the National Academy of Sciences},
   volume = {118},
   number = {12},
   pages = {e2022987118},
   ISSN = {0027-8424},
   year = {2021},
   DOI = {https://doi.org/10.1073/pnas.2022987118},
   type = {Journal Article}
}

@article{RN31,
   author = {Manna, Raj Kumar and Shklyaev, Oleg E and Balazs, Anna C},
   title = {Chemically driven multimodal locomotion of active, flexible sheets},
   journal = {Langmuir},
   volume = {39},
   number = {2},
   pages = {780-789},
   ISSN = {0743-7463},
   year = {2023},
   DOI = {https://doi.org/10.1021/acs.langmuir.2c02666},
   type = {Journal Article}
}

@article{RN32,
   author = {Song, Jiaqi and Shklyaev, Oleg E and Sapre, Aditya and Balazs, Anna C and Sen, Ayusman},
   title = {Self‐Propelling Macroscale Sheets Powered by Enzyme Pumps},
   journal = {Angewandte Chemie},
   volume = {136},
   number = {6},
   pages = {e202311556},
   ISSN = {0044-8249},
   year = {2024},
   DOI = {https://doi.org/10.1002/anie.202311556},
   type = {Journal Article}
}

@article{RN33,
   author = {Nakata, Satoshi and Tenno, Ryoichi and Deguchi, Ayano and Yamamoto, Hiroya and Hiraga, Yoshikazu and Izumi, Shunsuke},
   title = {Marangoni flow around a camphor disk regenerated by the interaction between camphor and sodium dodecyl sulfate molecules},
   journal = {Colloids and Surfaces A: Physicochemical and Engineering Aspects},
   volume = {466},
   pages = {40-44},
   ISSN = {0927-7757},
   year = {2015},
   DOI = {https://doi.org/10.1016/j.colsurfa.2014.10.041},
   type = {Journal Article}
}

@article{RN34,
   author = {Patiño, Tania and Arqué, Xavier and Mestre, Rafael and Palacios, Lucas and Sánchez, Samuel},
   title = {Fundamental Aspects of Enzyme-Powered Micro- and Nanoswimmers},
   journal = {Accounts of Chemical Research},
   volume = {51},
   year = {2018},
   DOI = {https://doi.org/10.1021/acs.accounts.8b00288},
   type = {Journal Article}
}

@article{RN35,
author = {Ma, Xing and Hortelão, Ana C. and Pati{\~n}o, Tania and Sánchez, Samuel},
title = {Enzyme Catalysis To Power Micro/Nanomachines},
journal = {ACS Nano},
volume = {10},
number = {10},
pages = {9111-9122},
year = {2016},
doi = {https://doi.org/10.1021/acsnano.6b04108},

}

@article{RN36,
   author = {Wang, Xiang and Li, Zhihao and Wang, Shuxu and Sano, Koki and Sun, Zhifang and Shao, Zhenhua and Takeishi, Asuka and Matsubara, Seishiro and Okumura, Dai and Sakai, Nobuyuki},
   title = {Mechanical nonreciprocity in a uniform composite material},
   journal = {Science},
   volume = {380},
   number = {6641},
   pages = {192-198},
   ISSN = {0036-8075},
   year = {2023},
  doi = {https://doi.org/10.1126/science.adf1206}, 
   type = {Journal Article}
}

@article{RN37,
   author = {Jiao, Dejin and Zhu, Qing Li and Li, Chen Yu and Zheng, Qiang and Wu, Zi Liang},
   title = {Programmable morphing hydrogels for soft actuators and robots: from structure designs to active functions},
   journal = {Accounts of Chemical Research},
   volume = {55},
   number = {11},
   pages = {1533-1545},
   ISSN = {0001-4842},
   year = {2022},
   doi = {https://doi.org/10.1021/acs.accounts.2c00046}, 
   type = {Journal Article}
}

@article{RN38,
   author = {Liu, Tianyi and Xie, Lei and Price, Cameron-Alexander Hurd and Liu, Jian and He, Qiang and Kong, Biao},
   title = {Controlled propulsion of micro/nanomotors: operational mechanisms, motion manipulation and potential biomedical applications},
   journal = {Chemical Society Reviews},
   volume = {51},
   number = {24},
   pages = {10083-10119},
   ISSN = {0306-0012},
   DOI = {10.1039/D2CS00432A},
   url = {http://dx.doi.org/10.1039/D2CS00432A},
   year = {2022},
   type = {Journal Article}
}

@article{RN39,
   author = {Serra-Casablancas, Meritxell and Di Carlo, Valerio and Esporrín-Ubieto, David and Prado-Morales, Carles and Bakenecker, Anna C. and Sánchez, Samuel},
   title = {Catalase-Powered Nanobots for Overcoming the Mucus Barrier},
   journal = {ACS Nano},
   volume = {18},
   number = {26},
   pages = {16701-16714},
   ISSN = {1936-0851},
   DOI = {10.1021/acsnano.4c01760},
   url = {https://doi.org/10.1021/acsnano.4c01760},
   year = {2024},
   type = {Journal Article}
}

@article{RN40,
   author = {Mathesh, Motilal and Sun, Jiawei and Wilson, Daniela A.},
   title = {Enzyme catalysis powered micro/nanomotors for biomedical applications},
   journal = {Journal of Materials Chemistry B},
   volume = {8},
   number = {33},
   pages = {7319-7334},
   ISSN = {2050-750X},
   DOI = {10.1039/D0TB01245A},
   url = {http://dx.doi.org/10.1039/D0TB01245A},
   year = {2020},
   type = {Journal Article}
}

@article{RN41,
   author = {Ghidiu, Michael and Lukatskaya, Maria R. and Zhao, Meng-Qiang and Gogotsi, Yury and Barsoum, Michel W.},
   title = {Conductive two-dimensional titanium carbide ‘clay’ with high volumetric capacitance},
   journal = {Nature},
   volume = {516},
   number = {7529},
   pages = {78-81},
   ISSN = {1476-4687},
   DOI = {https://doi.org/10.1038/nature13970},
   url = {https://doi.org/10.1038/nature13970},
   year = {2014},
   type = {Journal Article}
}

@article{RN42,
   author = {Reddy, Naveen Krishna and Clasen, Christian},
   title = {Self-propelling micro-disks},
   journal = {Korea-Australia Rheology Journal},
   volume = {26},
   number = {1},
   pages = {73-79},
   ISSN = {2093-7660},
   DOI = {https://doi.org/10.1007/s13367-014-0008-2},
   url = {https://doi.org/10.1007/s13367-014-0008-2},
   year = {2014},
   type = {Journal Article}
}

@article{RN43,
   author = {Valdez, Lyanne and Shum, Henry and Ortiz-Rivera, Isamar and Balazs, Anna C and Sen, Ayusman},
   title = {Solutal and thermal buoyancy effects in self-powered phosphatase micropumps},
   journal = {Soft Matter},
   volume = {13},
   number = {15},
   pages = {2800-2807},
   year = {2017},
   DOI = {https://doi.org/10.1039/C7SM00022G},
   type = {Journal Article}
}

@article{RN44,
   author = {Lai, Janice H and del Alamo, Juan C and Rodríguez-Rodríguez, Javier and Lasheras, Juan C},
   title = {The mechanics of the adhesive locomotion of terrestrial gastropods},
   journal = {Journal of Experimental Biology},
   volume = {213},
   number = {22},
   pages = {3920-3933},
   ISSN = {1477-9145},
   year = {2010},
   DOI = {https://doi.org/10.1242/jeb.046706},
   type = {Journal Article}
}

@article{RN45,
   author = {Budroni, MA and Upadhyay, Virat and Rongy, Laurence},
   title = {Making a simple A+ B→ C reaction oscillate by coupling to hydrodynamic effect},
   journal = {Physical review letters},
   volume = {122},
   number = {24},
   pages = {244502},
   ISSN = {0031-9007},
   year = {2019},
    DOI = {https://doi.org/10.1103/PhysRevLett.122.244502},
   type = {Journal Article}
}

@article{RN46,
   author = {Deng, Zixuan and Li, Kai and Priimagi, Arri and Zeng, Hao},
   title = {Light-steerable locomotion using zero-elastic-energy modes},
   journal = {Nature Materials},
   volume = {23},
   number = {12},
   pages = {1728-1735},
   ISSN = {1476-4660},
   DOI = {10.1038/s41563-024-02026-4},
   url = {https://doi.org/10.1038/s41563-024-02026-4},
   year = {2024},
   type = {Journal Article}
}

@article{RN47,
   author = {Zhu, Qing Li and Liu, Weixuan and Khoruzhenko, Olena and Breu, Josef and Hong, Wei and Zheng, Qiang and Wu, Zi Liang},
   title = {Animating hydrogel knotbots with topology-invoked self-regulation},
   journal = {Nature Communications},
   volume = {15},
   number = {1},
   pages = {300},
   ISSN = {2041-1723},
   DOI = {10.1038/s41467-023-44608-x},
   url = {https://doi.org/10.1038/s41467-023-44608-x},
   year = {2024},
   type = {Journal Article}
}

@article{RN48,
   author = {Lin, Han and Wang, Xingang and Yu, Luodan and Chen, Yu and Shi, Jianlin},
   title = {Two-Dimensional Ultrathin MXene Ceramic Nanosheets for Photothermal Conversion},
   journal = {Nano Letters},
   volume = {17},
   number = {1},
   pages = {384-391},
   note = {doi: 10.1021/acs.nanolett.6b04339},
   ISSN = {1530-6984},
   DOI = {10.1021/acs.nanolett.6b04339},
   url = {https://doi.org/10.1021/acs.nanolett.6b04339},
   year = {2017},
   type = {Journal Article}
}

@article{RN49,
   author = {Han, Xiaoxia and Huang, Ju and Lin, Han and Wang, Zhigang and Li, Pan and Chen, Yu},
   title = {2D Ultrathin MXene-Based Drug-Delivery Nanoplatform for Synergistic Photothermal Ablation and Chemotherapy of Cancer},
   journal = {Advanced Healthcare Materials},
   volume = {7},
   number = {9},
   pages = {1701394},
   ISSN = {2192-2640},
   DOI = {https://doi.org/10.1002/adhm.201701394},
   url = {https://advanced.onlinelibrary.wiley.com/doi/abs/10.1002/adhm.201701394},
   year = {2018},
   type = {Journal Article}
}

@article{RN50,
   author = {Caldarelli, Paolo and Chamolly, Alexander and Villedieu, Aurélien and Alegria-Prévot, Olinda and Phan, Carole and Gros, Jerome and Corson, Francis},
   title = {Self-organized tissue mechanics underlie embryonic regulation},
   journal = {Nature},
   volume = {633},
   number = {8031},
   pages = {887-894},
   ISSN = {1476-4687},
   DOI = {10.1038/s41586-024-07934-8},
   url = {https://doi.org/10.1038/s41586-024-07934-8},
   year = {2024},
   type = {Journal Article}
}

@article{RN51,
   author = {Hannezo, Edouard and Heisenberg, Carl-Philipp},
   title = {Mechanochemical feedback loops in development and disease},
   journal = {Cell},
   volume = {178},
   number = {1},
   pages = {12-25},
   ISSN = {0092-8674},
   year = {2019},
   DOI = {https://doi.org/10.1016/j.cell.2019.05.052},
   type = {Journal Article}
}

@article{RN52,
author = {Alhabeb, Mohamed and Maleski, Kathleen and Anasori, Babak and Lelyukh, Pavel and Clark, Leah and Sin, Saleesha and Gogotsi, Yury},
title = {Guidelines for Synthesis and Processing of Two-Dimensional Titanium Carbide (Ti3C2Tx MXene)},
journal = {Chemistry of Materials},
volume = {29},
number = {18},
pages = {7633-7644},
year = {2017},
doi = {https://doi.org/10.1021/acs.chemmater.7b02847},
}

@article{RN53,
   author = {Qiao, Qianqian and Wang, Jinyu and Long, Kai and Li, Linwei and Chen, Jiahao and Guo, Yuhao and Xu, Ziqiang and Kuang, Ying and Ji, Tianjiao and Li, Cao},
   title = {A cascaded enzyme system based on the catalase-like activity of Ti3C2Tx MXene nanosheets for the efficient combination cancer therapy},
   journal = {Nano Today},
   volume = {54},
   pages = {102059},
   ISSN = {1748-0132},
   year = {2024},
   doi = {https://doi.org/10.1016/j.nantod.2023.102059},
   type = {Journal Article}
}

@article{RN54,
   author = {Ling, Zheng and Ren, Chang E and Zhao, Meng-Qiang and Yang, Jian and Giammarco, James M and Qiu, Jieshan and Barsoum, Michel W and Gogotsi, Yury},
   title = {Flexible and conductive MXene films and nanocomposites with high capacitance},
   journal = {Proceedings of the National Academy of Sciences},
   volume = {111},
   number = {47},
   pages = {16676-16681},
   ISSN = {0027-8424},
   year = {2014},
   doi = {https://doi.org/10.1073/pnas.1414215111},
   type = {Journal Article}
}

@article{RN55,
   author = {Song, Jiaqi and Zhang, Jianhua and Lin, Jinwei and Shklyaev, Oleg E. and Shrestha, Shanid and Sapre, Aditya and Balazs, Anna C. and Sen, Ayusman},
   title = {Programming Fluid Motion Using Multi-Enzyme Micropump Systems},
   journal = {ACS Applied Materials \& Interfaces},
   volume = {16},
   number = {34},
   pages = {45660-45670},
   ISSN = {1944-8244},
   DOI = {10.1021/acsami.4c07865},
   url = {https://doi.org/10.1021/acsami.4c07865},
   year = {2024},
   type = {Journal Article}
}

\noindent LaTeX formats citations and references automatically using the bibliography records in your .bib file, which you can edit via the project menu. Use the cite command for an inline citation, e.g.  \cite{Hao:gidmaps:2014}.

For data citations of datasets uploaded to e.g. \emph{figshare}, please use the \verb|howpublished| option in the bib entry to specify the platform and the link, as in the \verb|Hao:gidmaps:2014| example in the sample bibliography file.

\section*{Acknowledgements}

This work was supported by the National Natural Science Foundation of China (grant no. 22120102001), the National Key Research and Development Program of China (Grant No. SQ2025YFE0100844), and the Innovation and Entrepreneurship Team of Jiangsu Province (grant no. JSSCTD202241). The work is sponsored by the interdisciplinary platform of science and engineering of the China University of Mining and Technology.

\section*{Author contributions statement}

H. W. and H. L. contributed equally to this work. The manuscript was written through the contributions of all authors. All authors have given approval to the final version of the manuscript.

\paragraph{Keywords:} Chemo-hydrodynamics, waves, Fluid-structure interaction, Active locomotion

\section*{Additional information}

\begin{figure}[ht]
\centering
\includegraphics[width=\linewidth]{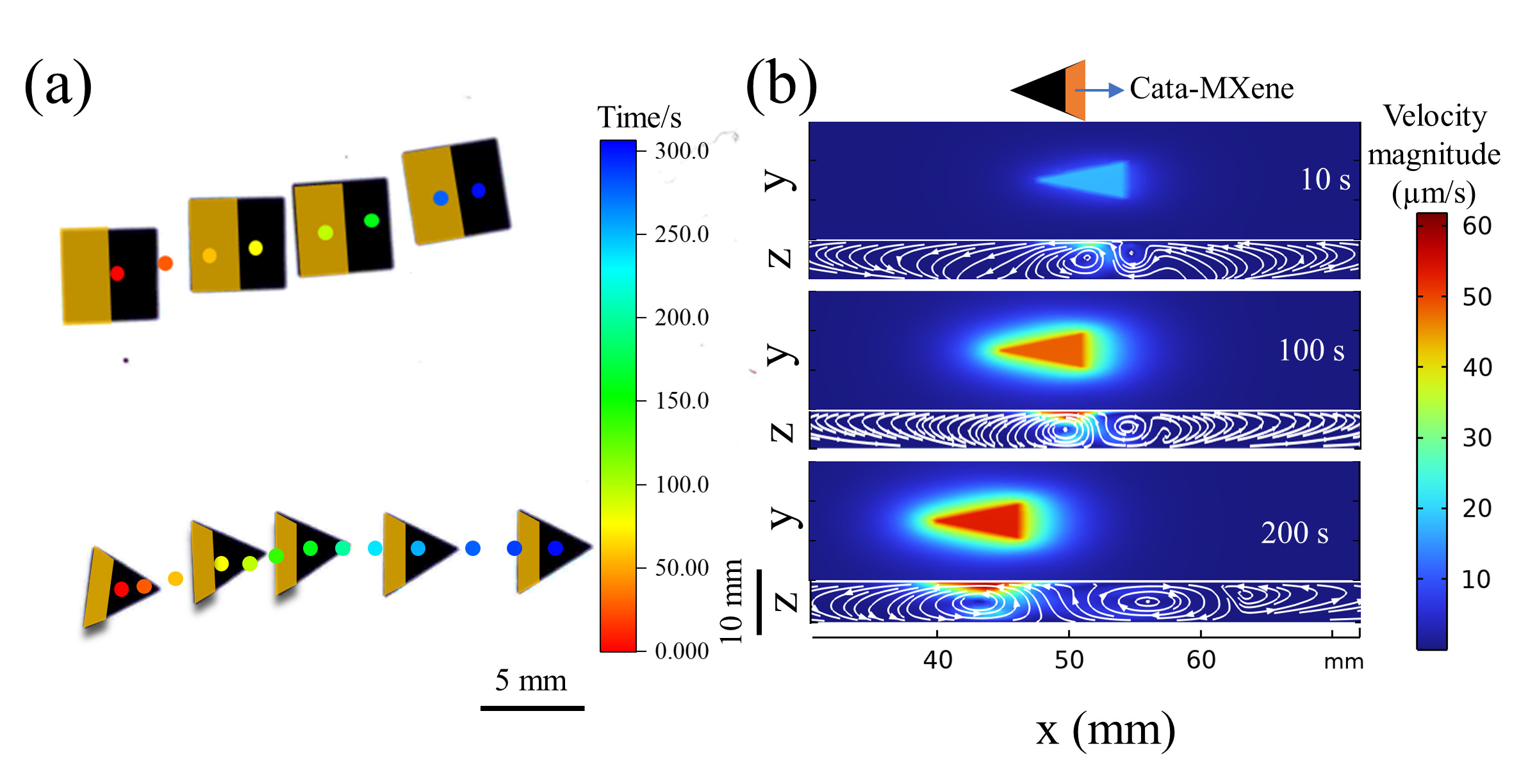}
\caption{linear motion of mirror-symmetric MXene sheets with one-side enzyme modification. (a) linear motion of square and triangular MXene sheets; the enzyme-coated region is shown in yellow. (b) Numerical simulation of the solutal buoyancy-driven directional motion of a triangular MXene sheet. The color encodes the total velocity magnitude. At each instant, both top-view (xy-plane) and side-view (xz-plane) contours are shown, and white curves represent streamlines.}
\label{}
\end{figure}

\begin{figure}[ht]
\centering
\includegraphics[width=\linewidth]{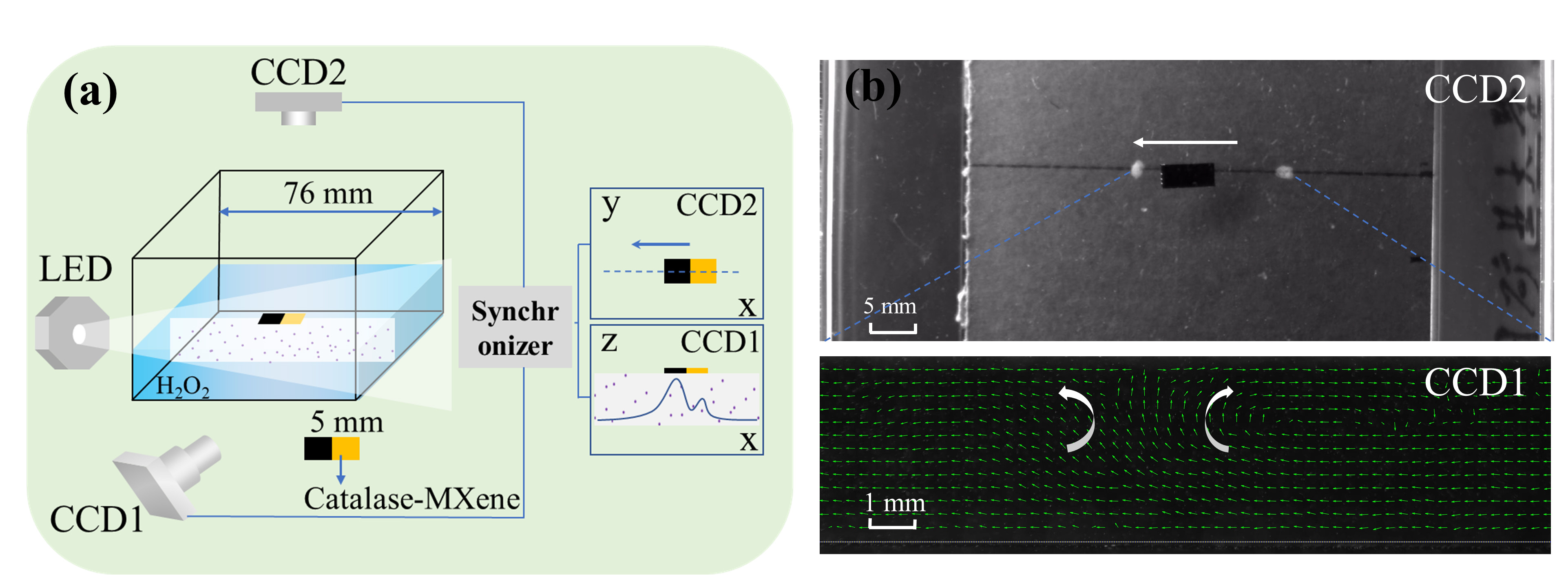}
\caption{Dual-view PIV system. (a) Schematic illustration of the experimental setup. (b) Dual-view imaging results: CCD1 captures the side-view PIV measurement of the flow field, and CCD2 presents the top-view motion of the active sheet at the air–liquid interface.}
\label{}
\end{figure}

\begin{figure}[ht]
\centering
\includegraphics[width=\linewidth]{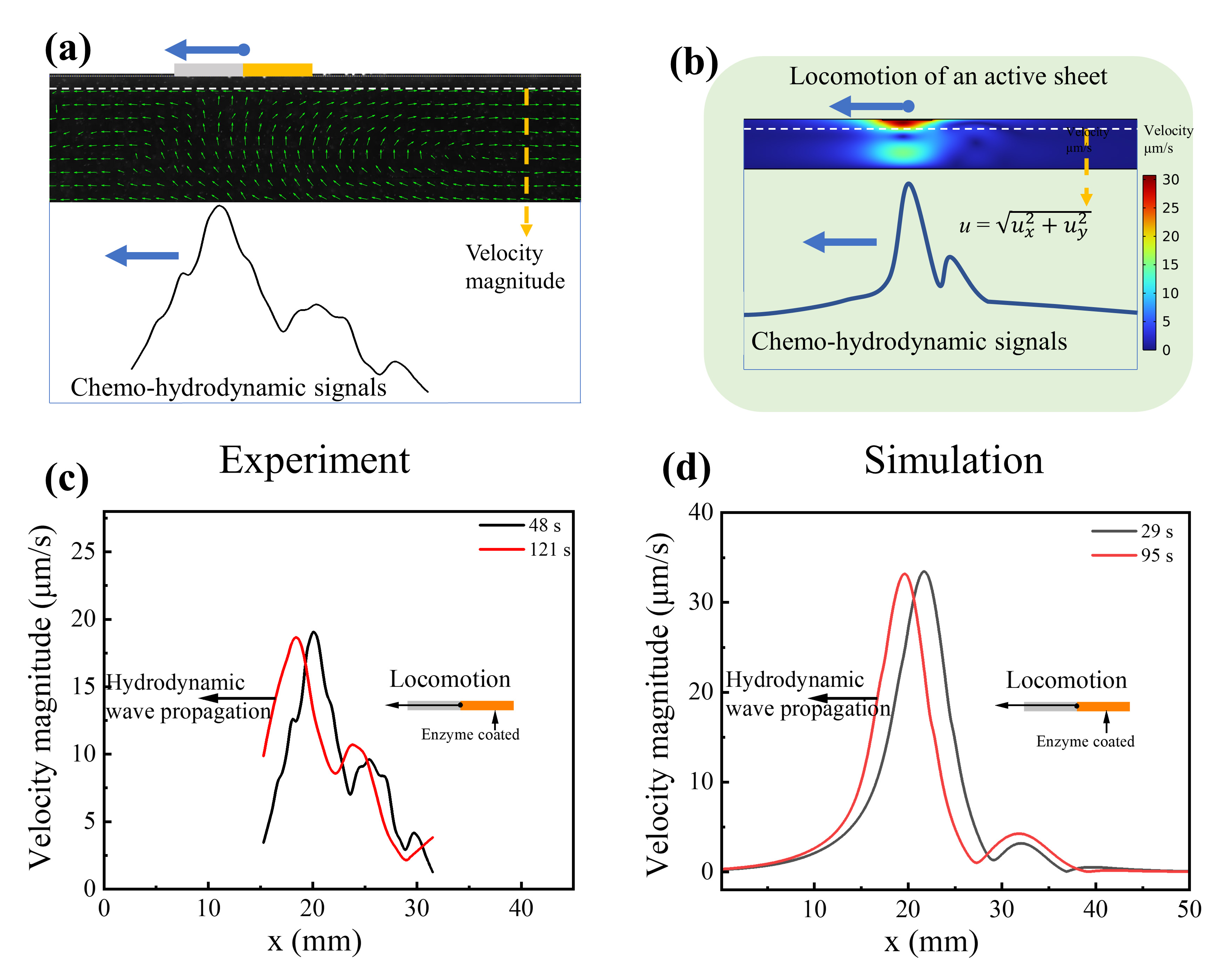}
\caption{Hydrodynamic wave propagation. (a) Hydrodynamic wave signal observed in the experiment at a substrate concentration of 10 mM and a measurement plane height of H = 4.8 mm. Green arrows indicate the local flow velocity direction. (b) Schematic of the hydrodynamic wave signal under the same conditions (10 mM, H = 4.8 mm). The color bar represents the velocity magnitude. (c, d) Time-sequence images of the wave signal from (c) experiment and (d) numerical simulation, demonstrating signal-driven locomotion.}
\label{}
\end{figure}

\begin{figure}[ht]
\centering
\includegraphics[width=\linewidth]{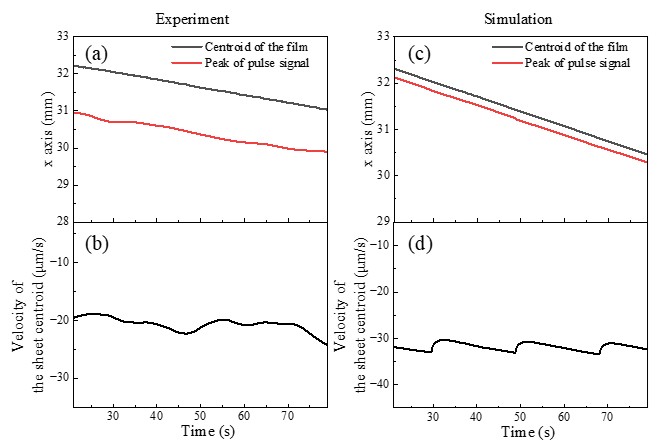}
\caption{Experimental and simulated chemo-hydrodynamic signals and the resulting active sheet motion. (a, b) Experimental data: displacement of the sheet centroid and propagation of the fluid wave, as well as the temporal variation of the sheet centroid speed (noise processing details in SI Figure S9a, b). (c, d) Simulation results: displacement of the sheet centroid and propagation of the fluid wave, along with the time-dependent sheet centroid speed.}
\label{}
\end{figure}

\begin{figure}[ht]
\centering
\includegraphics[width=\linewidth]{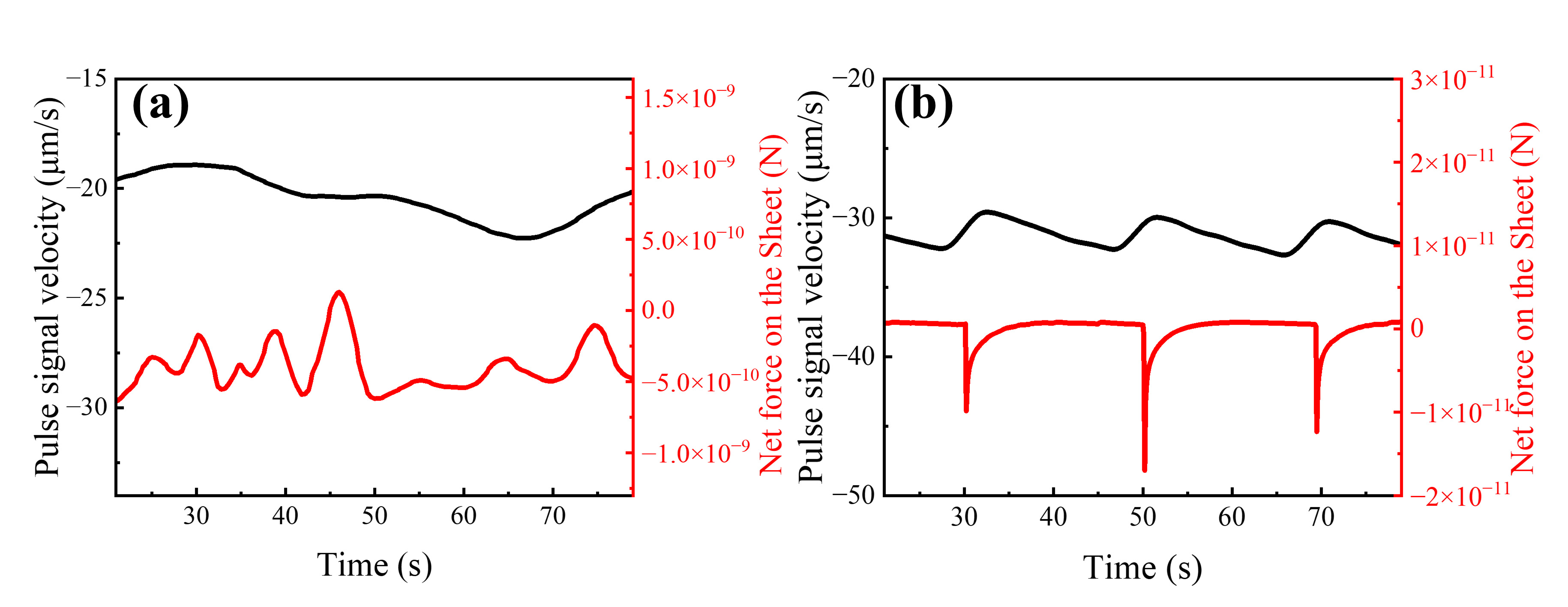}
\caption{Propulsion-analysis simulations of the active sheet. (a) Velocity of the chemo-hydrodynamic signal (black curve, left axis) and the corresponding net propulsion force acting on the sheet (red curve, right axis) as functions of time. (b) Corresponding simulation snapshot associated with panel (a).}
\label{}
\end{figure}

\clearpage

\section*{Supporting Information}

\subsection*{Section A: Experiment and characterization}

\subsubsection*{Preparation of asymmetric MXene sheets}

MXene materials have garnered significant attention since their discovery due to their exceptional hydrophilicity, high specific surface area, biocompatibility, electrical conductivity, and photothermal conversion properties, enabling broad applications in actuators and biomedical fields. As noted in the experimental section, MXene was prepared by in situ generation of hydrofluoric acid etching of Ti$_3$AlC$_2$ \cite{RN41}, followed by asymmetric deposition to obtain the actuation sheets used for the reaction. The aminated MXene nanosheets were crosslinked with catalase and assembled into a freestanding sheet via vacuum-assisted filtration (Figure S1b). The thickness of the sheets was precisely controlled by adjusting the volume of the colloidal solution. To ensure consistency, sheets with a thickness of approximately 5 $\mu$m (Figure S4c, SEM) were used for the motion studies. This design leverages the intrinsic properties of MXene and enzyme functionality, providing a promising platform for smart interfacial systems.

\begin{center}
\includegraphics[width=0.8\linewidth]{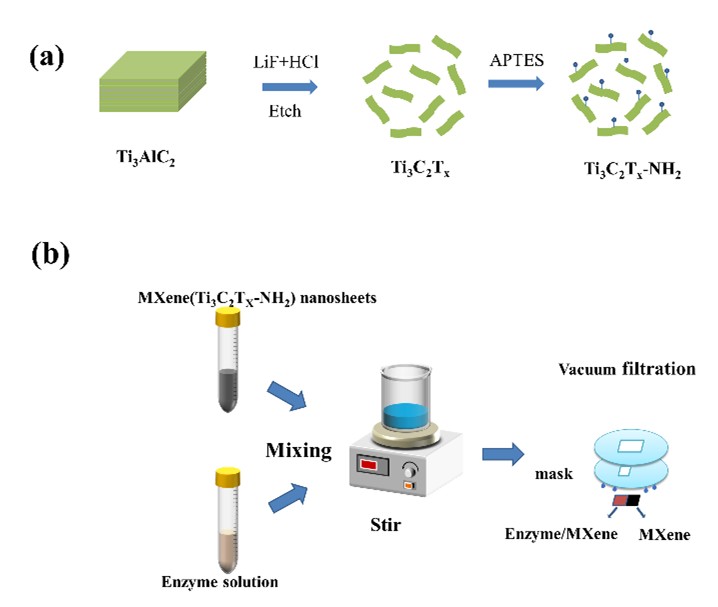}
\end{center}
\vspace{0.3em}
\noindent\textbf{Figure S1.} (a) Preparation of amino-functionalized MXene (Ti$_3$C$_2$T$_x$-NH$_2$). (b) Preparation of asymmetric MXene sheet by mask-assisted vacuum filtration.

\subsubsection*{Characterization of MXene nanosheets by SEM, AFM, and EDS spectroscopy}

The prepared MXene nanosheets form a stable colloidal dispersion (Fig.~S2a). SEM images (Figs.~S2c and S2d) reveal lateral dimensions on the micrometer scale, while AFM characterization indicates a thickness of approximately 2--3 nm.

\begin{center}
\includegraphics[width=\linewidth]{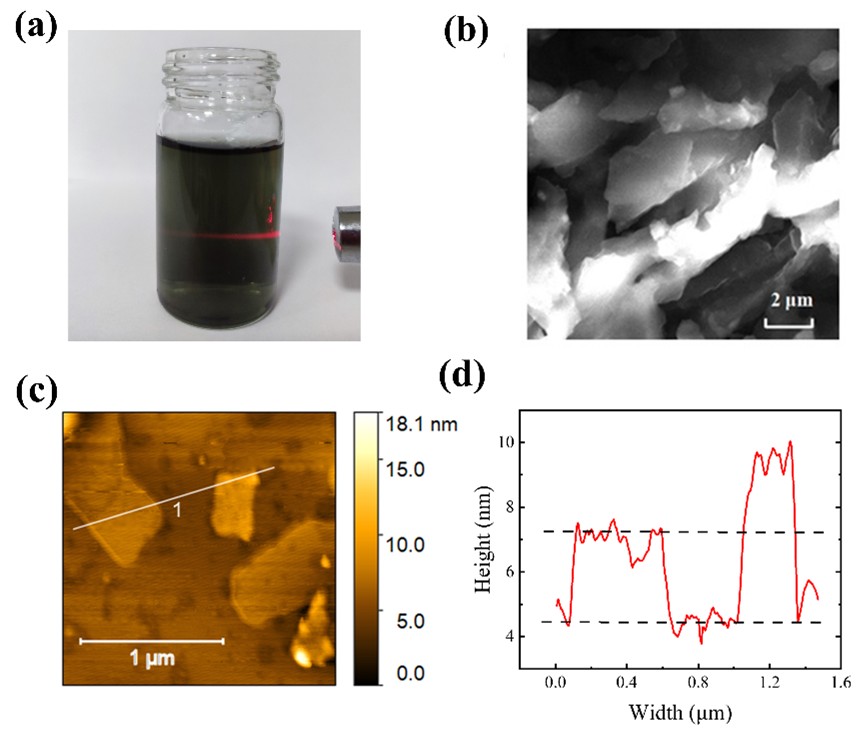}
\end{center}
\vspace{0.3em}
\noindent\textbf{Figure S2.} (a) Tyndall effect for the MXene colloidal solution. (b) SEM image of MXene nanosheets. (c) Surface topography images obtained by AFM. (d) Cross-sectional height profile along line 1 in (c), showing the thickness variation of the sheet.

\begin{center}
\includegraphics[width=\linewidth]{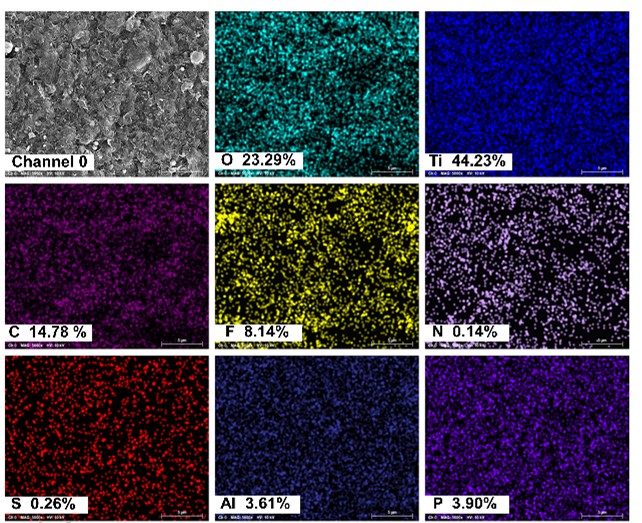}
\end{center}
\vspace{0.3em}
\noindent\textbf{Figure S3.} EDS elemental maps of the composite MXene sheet, showing the distribution of constituent elements. In these maps, colors represent the spatial distribution of the elements indicated in the lower left corner, brightness reflects the relative enrichment of each element, and the percentages denote the respective elemental compositions of the material.

\subsubsection*{Characterization of asymmetric MXene sheets by XRD, FTIR, and zeta potential analysis}

The as-prepared Ti$_3$C$_2$T$_x$ nanosheets were functionalized with aminopropyltriethoxysilane (APTES) to introduce --NH$_2$ groups, resulting in a positively charged surface (zeta potential shift from $-$30.3 to $+$21.6 mV, Figure S4d). X-ray diffraction (XRD) analysis confirmed the successful synthesis of Ti$_3$C$_2$T$_x$, as evidenced by the disappearance of the Ti$_3$AlC$_2$ (MAX phase) peak at $2\theta \approx 38^\circ$ (Figure S3a). The addition of PVA induced a leftward shift of the MXene (002) peak, indicative of increased interlayer spacing. Fourier-transform infrared (FT-IR) spectroscopy (Figure S3b) revealed characteristic peaks of Ti$_3$C$_2$T$_x$ at $\sim$3440 cm$^{-1}$ (O--H stretching vibration of surface-adsorbed water) and $\sim$1630 cm$^{-1}$ (O--H bending vibration). After APTES modification, the emergence of peaks at 2980--2880 cm$^{-1}$ corresponded to asymmetric C--H stretching vibrations in --CH$_3$ and --CH$_2$ groups. Distinct Si--O--Si and Si--OH asymmetric vibrations were observed at $\sim$1133 and 1050--1020 cm$^{-1}$, respectively. For the Ti$_3$C$_2$T$_x$-Catalase/PVA composite, catalase-specific peaks appeared at 1700 cm$^{-1}$ (C=O stretching), 1572 cm$^{-1}$ (amide II band), and 1388 cm$^{-1}$ (amide III band), confirming the successful immobilization of catalase on the MXene surface.

\begin{center}
\includegraphics[width=\linewidth]{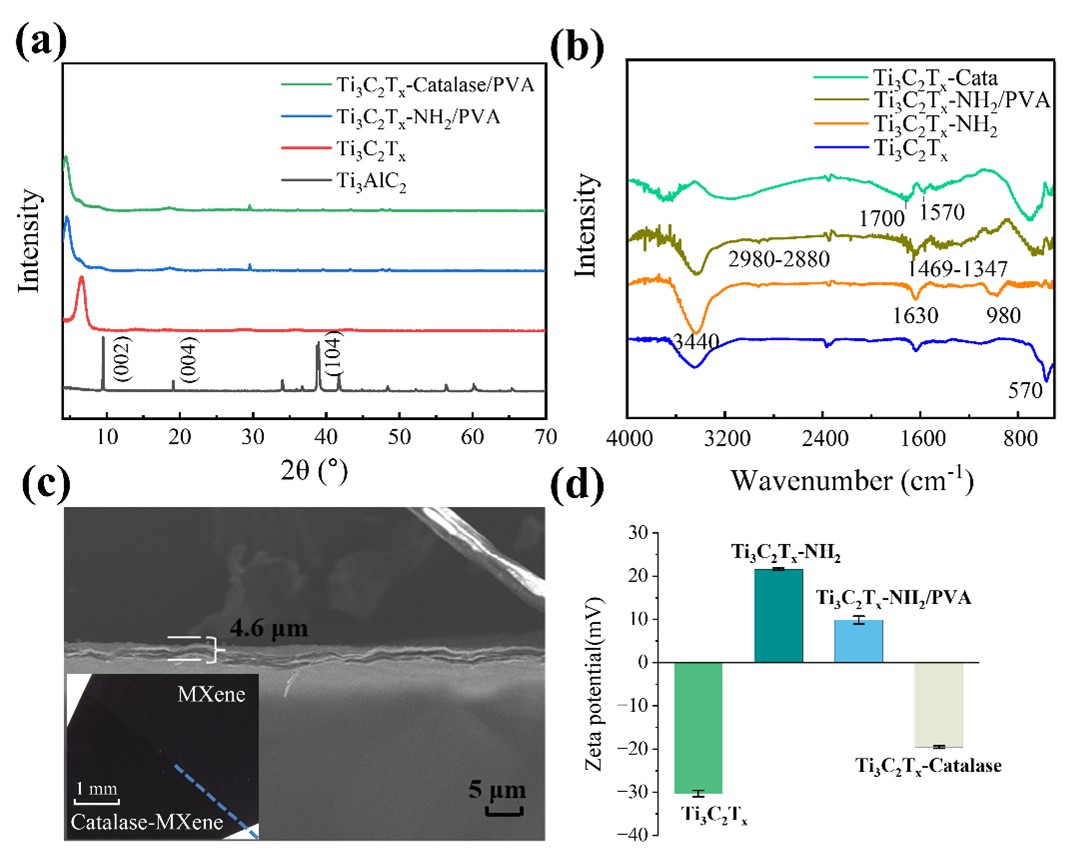}
\end{center}
\vspace{0.3em}
\noindent\textbf{Figure S4.} (a) XRD patterns of Ti$_3$AlC$_2$ and composite materials. (b) FTIR spectra. (c) Thickness of MXene composite sheet by SEM. The illustration shows the contact boundary between enzyme-loaded MXene sheet and bare MXene sheet. The darker regions correspond to the enzyme-loaded areas. (d) Zeta potential of composite MXene nanosheets.

\begin{center}
\includegraphics[width=0.7\linewidth]{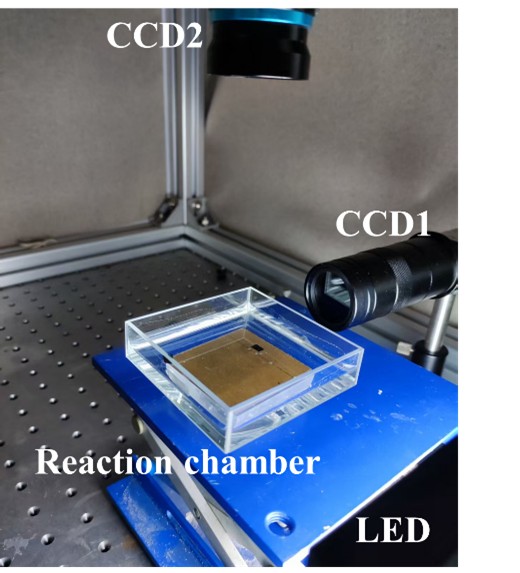}
\end{center}
\vspace{0.3em}
\noindent\textbf{Figure S5.} Photograph of the dual-view CCD particle image velocimetry (PIV) system. The two CCD cameras are connected to a synchronizer and operate synchronously under trigger signals. Both cameras are linked to a computer for image acquisition and recording.

\subsubsection*{Catalytic activity of active sheets and factors influencing locomotion}

\begin{center}
\includegraphics[width=\linewidth]{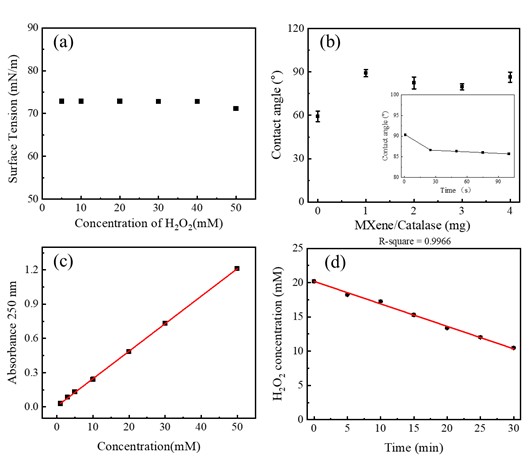}
\end{center}
\vspace{0.3em}
\noindent\textbf{Figure S6.} (a) Surface tension of solutions with varying hydrogen peroxide concentrations. Marangoni convection induced by the reaction is negligible. (b) Contact angle of the active membrane as a function of enzyme loading. The observed increase in hydrophobicity suggests enhanced binding between the enzyme and MXene hydrophilic groups. The inset shows the variation of the contact angle over time. (c) Immobilized-enzymatic activity measurements for catalase. Standardization curve of UV-vis absorbance at 250 nm versus hydrogen peroxide concentration. (d) Immobilized catalase activity is shown over time. Half of the 8 mm diameter circle was divided into eight parts and measured by an ultraviolet spectrophotometer with 10 mL of 20 mM hydrogen peroxide. The enzymatic reaction rate is 0.22 mM/min.

\subsubsection*{Linear locomotion and stationary states of active sheets}

\begin{center}
\includegraphics[width=\linewidth]{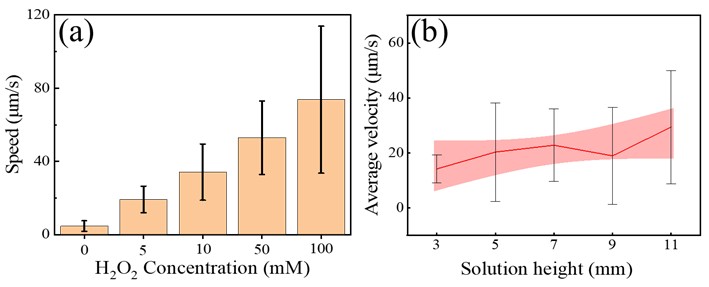}
\end{center}
\vspace{0.3em}
\noindent\textbf{Figure S7.} Factors affecting the linear motion of a rectangular sheet (2 $\times$ 5 mm, half-coated). (a) Mean velocity of the active sheet as a function of hydrogen peroxide concentration. (b) Mean velocity of the sheet's linear motion as a function of the liquid-layer height. The red shading denotes the confidence band.

\begin{center}
\includegraphics[width=0.7\linewidth]{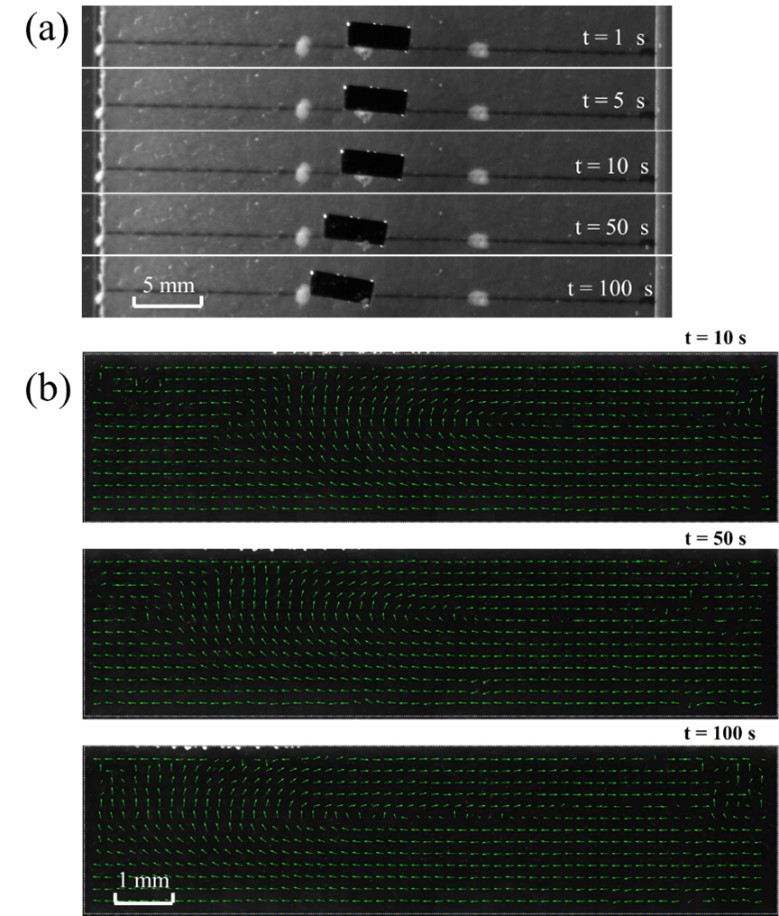}
\end{center}
\vspace{0.3em}
\noindent\textbf{Figure S8.} Time-sequence images captured by the dual-view PIV system. (a) Macroscopic linear motion of the sheet recorded by CCD2 (top view). (b) Side-view flow field recorded by CCD1 and processed with PIVlab; green arrows indicate the local flow velocity direction.

\begin{center}
\includegraphics[width=\linewidth]{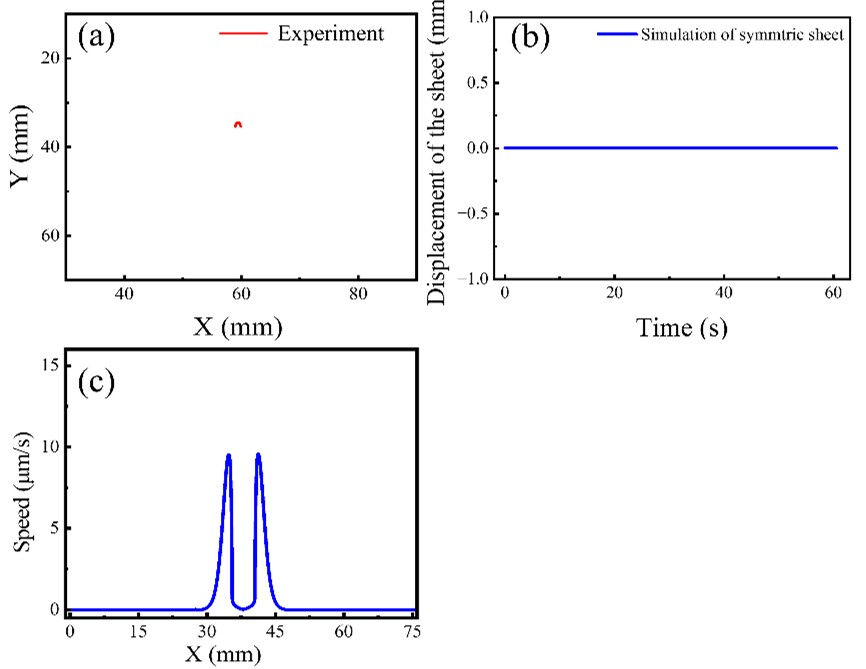}
\end{center}
\vspace{0.3em}
\noindent\textbf{Figure S9.} Motion of a symmetrically coated sheet. (a) Experimental trajectories (top view). (b) Simulated sheet displacement over time. (c) Symmetric distribution of velocity signals at the initial position of the sheet.

\subsubsection*{Wave signals at different solution heights}

\begin{center}
\includegraphics[width=\linewidth]{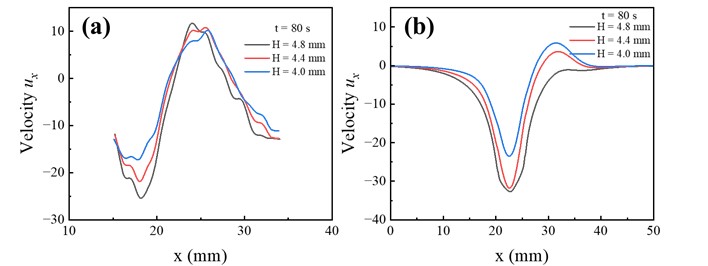}
\end{center}
\vspace{0.3em}
\noindent\textbf{Figure S10.} Experimental and simulated changes in waves (velocity $u_x$) around the active membrane with height. (a) Experimental observation. (b) Simulation results.

\subsubsection*{Noise processing during locomotion}

\begin{center}
\includegraphics[width=\linewidth]{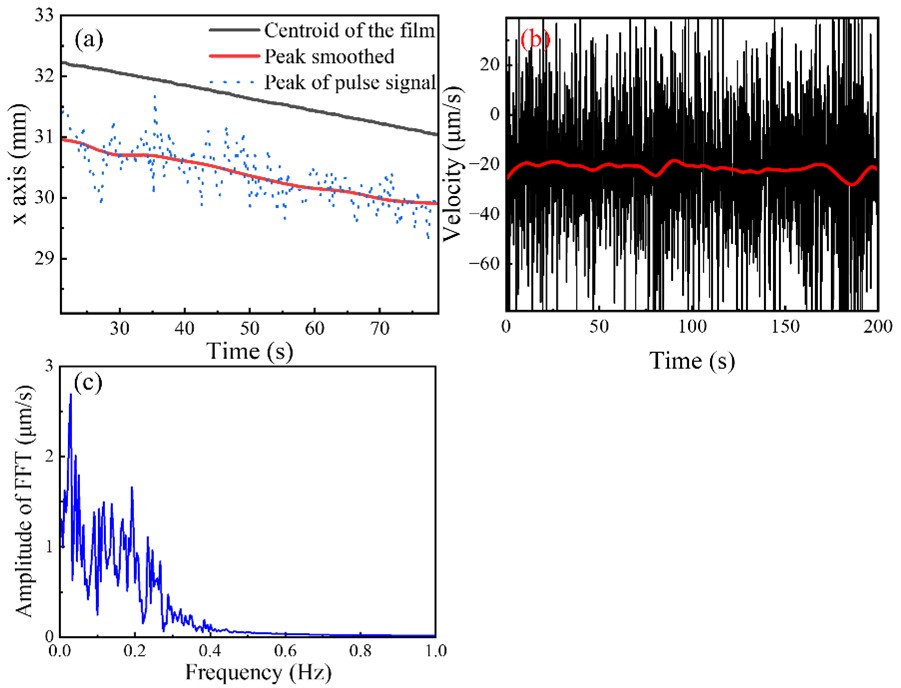}
\end{center}
\vspace{0.3em}
\noindent\textbf{Figure S11.} Experimental analysis for active sheet locomotion dynamics. (a) Distance difference between the sheet centroid and the lowess-smoothed signal (blue line). The raw wave peaks are shown as red dots. (b) Speed oscillations of the active sheet's centroid. To ensure frequency consistency and mitigate environmental noise, the same lowess smoothing method with an identical window width was applied. The resultant smoothed data (red line) is presented in Figure 4a (main text). (c) Fourier spectral analysis of the detrended centroid velocity shown in (b). Due to the relatively high noise level in the experimental system, a fourth-order Butterworth low-pass filter (cutoff frequency $f_c = 0.2$ Hz) was applied using zero-phase filtering to suppress high-frequency noise while preserving the dominant low-frequency dynamics. The resulting single-sided amplitude spectrum exhibits a pronounced peak in the low-frequency regime, suggesting the existence of a dominant low-frequency component in the velocity dynamics.

\subsection*{Section B: Numerical simulation}

The multiphysics 3D model was built in COMSOL Multiphysics, coupling fluid flow, chemical species transport, and structural mechanics. The fluid flow was governed by the Laminar Flow interface; the transport and reaction of chemical species were described by the Transport of Diluted Species interface; and the mechanical response of the asymmetric MXene membrane was accounted for via the Solid Mechanics interface. These physics were integrated using the Fluid--Structure Interaction (FSI) multiphysics interface to achieve bidirectional coupling.

The computational domain corresponds to the experimental chamber, a rectangular prism measuring 76 mm $\times$ 76 mm $\times$ 5 mm. The MXene sheet is represented by a rectangular solid domain of 5 mm $\times$ 2 mm $\times$ 0.1 mm that floats at the liquid surface, with only its bottom surface in contact with the liquid (Figure S11a). The analogous triangular sheet geometry is shown in Figure S11b.

Boundary conditions. The entire bottom face of the MXene sheet forms the fluid--structure interface and was set as a no-slip wall for the fluid. A spatially restricted reaction was imposed on the right half of this bottom face (indicated by the green segment in Figure S11a, green surface in Figure S11b illustration), where the normal species flux was prescribed according to Equations 4--6 in the main text; the left half of the bottom face was assigned zero species flux. All external boundaries of the fluid domain (the bottom and side walls of the reaction cell) were treated as no-slip walls with zero species flux. The top boundary of the fluid, representing the air--liquid interface outside the sheet, was assumed to remain flat and was modeled as a free-slip wall with no species flux. It is worth noting that the resultant force on the sheet in the simulation is calculated based on the entire bottom surface corresponding to the green regions in Figures S12a and S12b.

To account for the confinement of the sheet at the air--liquid interface, the normal mesh displacement on the bottom face of the flake was constrained to zero, thereby restricting the solid motion to the horizontal ($x$) direction; displacement in the vertical ($z$) direction was neglected.

Initial conditions. The fluid was initially quiescent with a uniform species concentration of $c_0 = 10$ mM and a velocity field of $u_0 = 0$ $\mu$m/s. The MXene sheet was initially at rest at its prescribed starting position.

\begin{center}
\includegraphics[width=\linewidth]{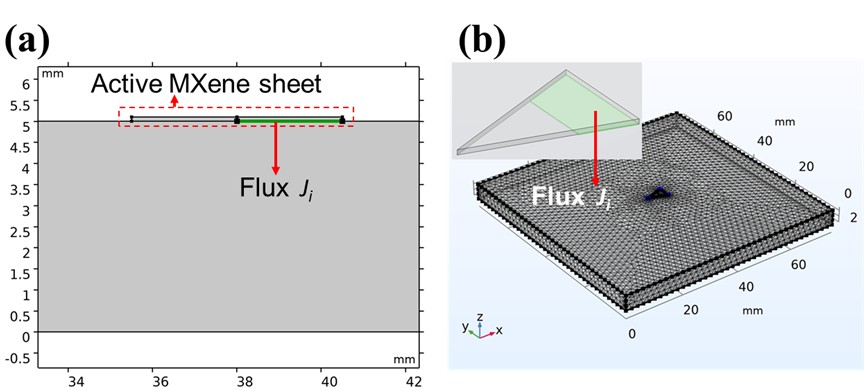}
\end{center}
\vspace{0.3em}
\noindent\textbf{Figure S12.} COMSOL mesh settings and boundary conditions. (a) Cross-section of the 3D model in the $xz$ plane. The MXene sheet is outlined by the red dashed box, and the green contact line on the right half of the bottom surface indicates the catalase-modified region. (b) 3D model of an isosceles triangular sheet with coarser mesh. Inset: the green area on the bottom surface denotes the catalase-loaded region.

The Transport of Diluted Species module governed the hydrogen peroxide decomposition kinetics, incorporating Michaelis-Menten enzymatic parameters from the literature \cite{RN30,RN32}: catalytic rate constant $k_{\text{cat}}~(s^{-1}) = 2.12 \times 10^{5}$, $n = 4$, Michaelis constant $K_m = 0.093$ M. Considering the catalase specific activity (3000 U mg$^{-1}$) and taking the total loaded activity on the sheet as approximately 3000 U over an area $S = 5$ mm$^2$, the surface enzyme concentration was estimated as $E = 3000\,U/(S \cdot n \cdot k_{\text{cat}}) \approx 1.18 \times 10^{-5}$ mol m$^{-2}$. Other parameters are detailed in Table S1 \cite{RN30}.

\begin{center}
\includegraphics[width=\linewidth]{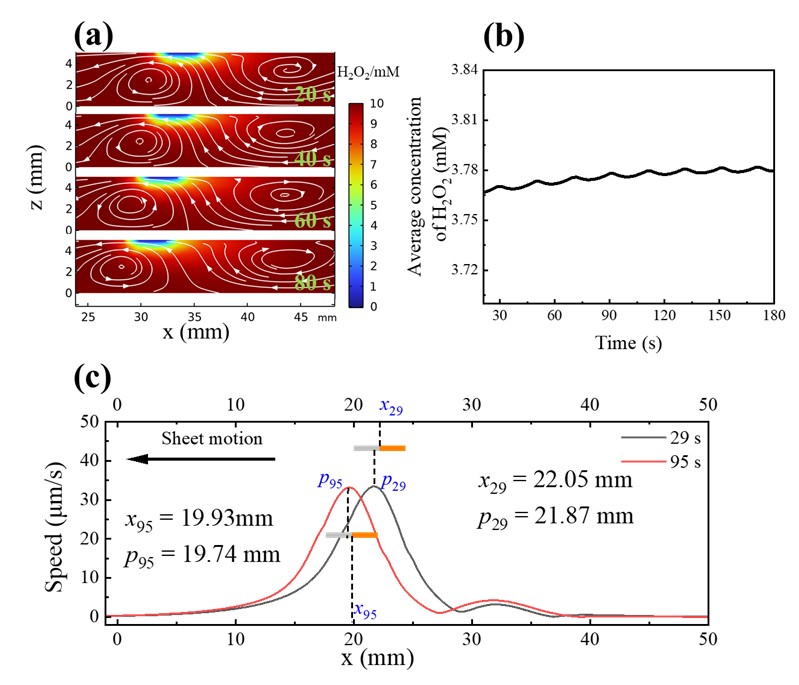}
\end{center}
\vspace{0.3em}
\noindent\textbf{Figure S13.} Detailed simulations of wave signals and the active sheet during motion. (a) Spatiotemporal map of the hydrogen peroxide concentration during the motion of the active sheet, the white lines represent the streamlines. (b) Temporal evolution of the average hydrogen peroxide (H$_2$O$_2$) concentration at the bottom of the active sheet. (c) Velocity curves of fluid signal at different times along the horizontal line $H = 4.8$ mm during its motion. $x_{29}$ and $x_{95}$ denote the center points of the sheet, and $p_{29}$ and $p_{95}$ denote the peak locations of waves at 29 s and 95 s, respectively.

\subsubsection*{Concentration dynamics with locomotion}

To further resolve the concentration dynamics accompanying the active sheet motion, we constructed a space--time plot of the H$_2$O$_2$ concentration evolution (Figure S13a). The color encodes the local H$_2$O$_2$ concentration, with an initial bulk concentration of 10 mM. As shown, the concentration gradient generated by the catalase-coated region of the sheet translates with the sheet as it moves. Adopting the sheet's own reference frame, we then recorded the H$_2$O$_2$ concentration at the bottom surface of the moving sheet at each instant, yielding the time trace in Figure S13b. This sampling dynamic strategy differs from earlier studies in which variates such as flow velocity or concentration were monitored at a fixed location over time\cite{RN45}. The temporal variation in Figure S13b reveals that, as the sheet enters fresh regions, the local H$_2$O$_2$ concentration it consumes fluctuates because of diffusional replenishment. This figure provides strong evidence that the velocity fluctuations originate from concentration variations that modulate the driving force.

\begin{center}
\textbf{Table S1.} Parameters of COMSOL simulation.
\vspace{0.3em}
\begin{tabular}{lccc}
\hline
\textbf{Parameters} & \textbf{Catalase} & \textbf{H$_2$O$_2$} & \textbf{H$_2$O} \\
\hline
$k_{\text{cat}}$ (s$^{-1}$)        & $2.12\times 10^{5}$ & --    & --   \\
$K_{\text{m}}$ (M)                & 0.093               & --    & --   \\
$E$ (M)                           & $1.18\times 10^{-5}$& --    & --   \\
$\beta$ (M$^{-1}$)                & --                  & 0.01056 & -- \\
$D$ (m$^2$/s)                     & --                  & $1.43\times 10^{-9}$ & $1\times 10^{-9}$ \\
$\rho$ (g/cm$^3$)                 & --                  & 1.45  & 1    \\
\hline
\end{tabular}
\end{center}

\end{document}